\documentclass[apj,twocolumn]{emulateapj}
\usepackage{apjfonts}

\newcommand{\beq}{\begin{equation}}
\newcommand{\eeq}{\end{equation}}
\newcommand{\beqn}{\begin{eqnarray}}
\newcommand{\eeqn}{\end{eqnarray}}

\newcommand{\AU}{{\rm AU}} 

\newcommand{\kB}{k_{\rm B}}

\newcommand{\eqref}[1]{(\ref{#1})}

\newcommand{\dfrac}[2]{ {\displaystyle\frac{#1}{#2}} }
\newcommand{\pfrac}[2]{ \Bigl(\dfrac{#1}{#2}\Bigr) }
\newcommand{\eps}{\epsilon}
\renewcommand{\leq}{\leqslant}
\renewcommand{\geq}{\geqslant}

\defcitealias{O09}{O09}
\defcitealias{OTS09}{OTS09}
\defcitealias{OTTS11a}{Paper I}

\shorttitle{ELECTROSTATIC BARRIER AGAINST DUST GROWTH. II}
\shortauthors{OKUZUMI ET AL.}

\begin{document}
\title{Electrostatic Barrier Against Dust Growth in Protoplanetary Disks. \\II. 
Measuring the Size of the ``Frozen'' Zone}
\author{Satoshi Okuzumi\altaffilmark{1,2}, Hidekazu Tanaka\altaffilmark{3},
Taku Takeuchi\altaffilmark{3,4}, and Masa-aki Sakagami\altaffilmark{1}}
\email{okuzumi@nagoya-u.jp}
\altaffiltext{1}{Graduate School of Human and Environmental Studies, Kyoto University,
Kyoto 606-8501, Japan}
\altaffiltext{2}{Department of Physics, Nagoya University, Nagoya, Aichi 464-8602, Japan}
\altaffiltext{3}{Institute of Low Temperature Science, Hokkaido University, Sapporo 060-0819, Japan}
\altaffiltext{4}{Department of Earth and Planetary Sciences, Tokyo Institute of Technology, Tokyo 152-8551, Japan}
\begin{abstract}
Coagulation of submicron-sized dust grains into porous aggregates is the initial step of dust evolution in protoplanetary disks. Recently, it has been pointed out that negative charging of dust in the weakly ionized disks could significantly slow down the coagulation process. In this paper, we apply the growth criteria obtained in Paper I to finding out a location (``frozen'' zone) where the charging stalls dust growth at the fractal growth stage. For low-turbulence disks, we find that the frozen zone can cover the major part of the disks at a few to 100 AU from the central star. The maximum mass of the aggregates is approximately $10^{-7}~{\rm g}$ at 1 AU and as small as a few monomer masses at 100 AU. Strong turbulence can significantly reduce the size of the frozen zone, but such turbulence will cause the fragmentation of macroscopic aggregates at later stages. We examine a possibility that complete freezeout of dust evolution in low-turbulence disks could be prevented by global transport of dust in the disks. Our simple estimation shows that global dust transport can lead to the supply of macroscopic aggregates and the removal of frozen aggregates on a timescale of $10^6$ yr. This overturns the usual understanding that tiny dust particles get depleted on much shorter timescales unless collisional fragmentation is effective. The frozen zone together with global dust transport might explain ``slow'' ($\sim 10^6$ yr) dust evolution suggested by infrared observation of T Tauri stars and by radioactive dating of chondrites.
\end{abstract}
\maketitle

\section{Introduction}
Growth of submicron-sized interstellar dust grains into kilometer-sized planetesimals 
is the first step towards the formation of terrestrial planets and the cores of gas giants 
in protoplanetary disks \citep{Mizuno80,Pollack+96}.
The formation of planetesimals is believed to involve the following stages.
(1) Initially submicron-sized particles coagulate into larger but highly porous, fractal aggregates 
through Brownian motion and differential settling towards the midplane of the disk 
\citep{WB98,Blum+98,KPH99}.
(2)  As the aggregates grow to ``macroscopic'' (mm to cm) sizes, 
the collisional energy becomes high enough to cause the compaction of the aggregates 
\citep{Blum04,SWT08,PD09}.
(3) The compaction cause the decrease in the gas drag force acting on the aggregates,
allowing them to concentrate in the midplane of the disk \citep{Safronov69,GW73},
the center of vortices \citep{BS95}, and turbulent eddies \citep{Johansen+07}.
(4) Planetesimals may form within such dense regions through gravitational instability 
\citep{Safronov69,GW73} or through further collisional growth \citep{WC93,W95}.

However, a number of obstacles, often called ``barriers,'' have been pointed out 
against the above processes.
As the collisional compaction proceeds,
the motion of aggregates relative to the gas becomes faster and faster 
due to radial drift \citep{W77} and random motion induced by turbulence \citep{Voelk+80}.
The collision velocity can exceed $10~{\rm m~s^{-1}}$ even without turbulence \citep{W77},
but it is uncertain whether such high-speed collisions lead to the sticking 
or fragmentation of the aggregates 
\citep[the ``fragmentation barrier'';][]{BW08, Wada+09,TW09,Guettler+10}.
Furthermore, recent laboratory experiments suggest that the compaction itself 
causes reduction of sticking efficiency and bouncing of colliding aggregates \citep{BW08, Guettler+10}.
This may halt dust growth before the aggregates reach the fragmentation barrier 
\citep[the ``bouncing barrier'';][]{Zsom+10}.
Even without fragmentation,
the radial drift can result in the lost of dust particles from the entire part of the disk 
unless their growth proceeds very rapidly \citep[the "radial drift barrier"; e.g.,][]{BDH08}.
Besides, turbulence does not necessarily promote the local concentration of dust, 
because it can efficiently diffuse small particles \citep{WC93}.

On the other hand, astronomical as well as meteoritic observations
suggest that some barriers against dust evolution may {\it do} exist in protoplanetary disks.
Mid-infrared excess observed for classical T Tauri stars \citep[e.g.,][]{Furlan+06}
implies a certain amount of small dust grains retained in the inner parts of 
their circumstellar disks over a million years.
This feature cannot be explained by simple coagulation theory 
assuming perfect sticking efficiency \citep{DD05}.
In addition, radioisotope dating of the most primitive chondrites 
supports that the formation of chondrules began 
at least a million years after the formation of the Solar nebula  \citep[e.g.][]{Kita+00,Kita+05}.
This also suggests that the dust growth process towards planetesimals 
is ``inefficient'' \citep{CHS08}.
Thus, for better modeling of dust evolution in protoplanetary disks,
knowledge on what prolongs it is as important as knowledge on what promotes it.

Recently, one of the authors has pointed out another kind of barriers,
namely, an electrostatic barrier due to dust charging \citep[][hereafter O09]{O09}.
Protoplanetary disks are expected to be weakly ionized 
by a various kinds of high-energy sources, such as cosmic rays \citep{UN81}
and X-rays from the central star \citep{GNI97,IG99}.
As is known in plasma physics  \citep{SM02},
dust particles negatively charge in an ionized gas
because free electrons hit to the particles more frequently than ions.
This ``asymmetric'' charging implies possible existence of an electrostatic barrier against dust growth,
but this effect has been ignored in the context of dust growth in protoplanetary disks. 
\citetalias{O09} estimated how strongly the charging affects the collisional cross section
between aggregates properly taking into account the weak ionization of the disks.
The result shows that the collisional cross section can be strongly suppressed 
{\it before} the collisional compaction becomes effective.
This is in clear contrast to the previously known barriers which act 
{\it after} the onset of the compaction.

\begin{figure}
\plotone{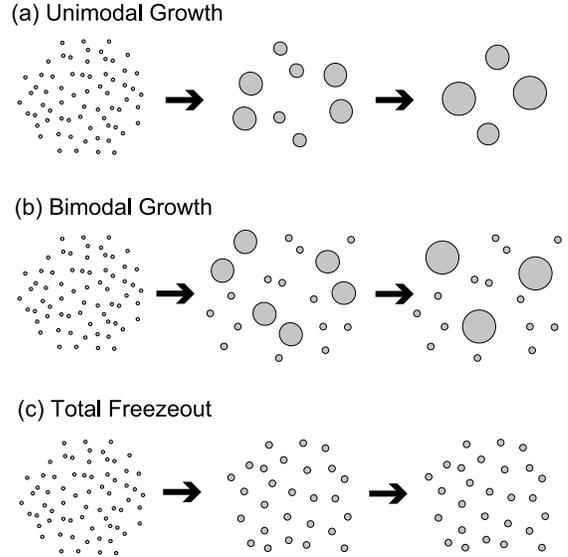}
\caption{
Schematic illustration showing 
three outcomes of the collisional growth of fractal dust aggregates charging 
in a weakly ionized gas:
(a) unimodal growth, (b) bimodal growth,
and (c) total freezeout.
In the unimodal growth, dust aggregates grow with a relatively narrow size distribution.
In the bimodal growth, a certain number of aggregates 
stop growing due to the electrostatic repulsion 
and the rest continue growing by colliding with each other.
In the total freezeout, all aggregates stop growing with a nearly monodisperse distribution.
See \citetalias{OTTS11a} for details.
}
\label{fig:outcome}
\end{figure}

However, the simple estimate by \citetalias{O09} 
assumed that dust aggregates grow with a narrow size distribution.
In reality, size distribution is determined as a result of the coagulation process,
and it has been unclear how the distribution evolves when the charging is taken into account.
The number of small aggregates is particularly important because 
it determines the ionization state of the gas and hence
the charge state of {\it all} aggregates \citepalias{O09}.
To address this issue, in our preceding paper \citep[][hereafter \citetalias{OTTS11a}]{OTTS11a},
we have numerically simulated how the size distribution 
evolves in the presence of the electrostatic repulsion
properly taking into account ionization balance in the gas--dust mixture \citepalias{O09} 
and porosity evolution due to low-velocity sticking
 \citep[][hereafter \citetalias{OTS09}]{OTS09}.
We find that the outcomes can be classified into three types: 
\begin{itemize}
\item[(a)] {\it Unimodal growth.} 
If the electrostatic repulsion is negligibly weak,  
aggregates simply grow with a relatively narrow size distribution (Figure~\ref{fig:outcome}(a)).
We refer to this growth mode as the ``unimodal growth.''

\item[(b)] {\it Bimodal growth.} 
If the electrostatic repulsion is strong but 
nonthermal motion (e.g., vertical sedimentation and turbulence) dominates aggregate collision,
some aggregates stop growing at a certain size
while the rest continue growing by colliding with each other (Figure~\ref{fig:outcome}(b)).
We call this the ``bimodal growth.''
In this mode, growing aggregates dominate the total dust mass, 
but their negative charges are suppressed by the non-growing (``frozen'') aggregates. 
Interestingly, in some cases, the presence of the frozen aggregates 
is even required for the larger aggregates to continue growing.
The bimodal growth thus demonstrates the importance of the dust size distribution.

\item[(c)] {\it Total freezeout.} 
If the electrostatic repulsion is strong and thermal (Brownian) 
motion dominates aggregate collision, all aggregates stop growing at a certain size
(Figure~\ref{fig:outcome}(c)).
We refer to this type of growth mode as the ``total freezeout.''
\end{itemize}
We have also obtained a set of simple criteria for which of the outcomes
is realized under given conditions.
These criteria allow us to predict how the initial fractal growth 
proceeds at different locations in protoplanetary disks.

In this paper, we use the growth criteria obtained in \citetalias{OTTS11a}
to map a region where local fractal dust growth ends up with the total freezeout,
to which we will refer as the ``{\it frozen zone}.''
This is the first step towards comprehensive modeling of dust evolution in protoplanetary disks
including dust charging together with 
collisional compaction
\citep{SWT08}, radial drift \citep{W77,BDH08}, bouncing \citep{Zsom+10}, 
and fragmentation \citep{BDH08,BDB09}. 
In Section~2, we briefly summarize the analysis done in \citetalias{OTTS11a}
and present the growth criteria.
The protoplanetary disk model used in this paper is described in Section 3,
and the main results are presented in Section 4.
In Section 5, we discuss potential mechanisms that could 
prevent dust growth in the frozen zone from completely frozen.
A summary of this paper is presented in Section~6. 

\section{Coagulation of Fractal Dust Aggregates in a Weakly Ionized Gas}
In this section, we outline the analysis done in \citetalias{OTTS11a}
and introduce several quantities to write down the growth criteria.

We focus on the first stage of dust evolution in protoplanetary disks
where submicron-sized dust particles (``monomers'') grow into fractal aggregates.
For simplicity, we assume that the size of the monomers is equal and treat it as a free parameter.
The assumption of fractal growth is valid only when the collision energy is low and 
both compaction and fragmentation is negligible \citep{DT97,SWT08}.
The limitation of this assumption will be shown in Section 4.1.
Note that most previous studies on dust coagulation ignored fractal evolution 
and treated aggregates as compact spheres \citep[e.g.,][]{NNH81,THI05,BDH08}.
However, as we will see in Section 5.1, it is critical to properly take into account fractal evolution 
when analyzing the electrostatic barrier.

Furthermore, we assume that the dust growth proceeds locally.
In laminar disks, this assumption is valid as long as dust grows into fractal aggregates,
because the timescales of vertical settling and radial drift are much 
longer than that of  local growth.
In turbulent disks, dust can be globally transported on a short timescale by turbulent mixing, 
but we ignore this effect in Sections 2--4.
Effects of turbulent mixing as well as radial drift will be discussed in Section 5.2.

The collision velocity between aggregates is assumed to be 
driven by thermal (Brownian) motion and nonthermal ``differential drift.''
Here, differential drift refers to dust motion whose relative velocity 
has of the form $\Delta u_D = g|\tau_{f,1} -\tau_{f,2}|$,
where $\tau_{f,1}$ and $\tau_{f,2}$ are the stopping times of colliding aggregates 
and $g$ is the effective acceleration inducing the relative motion.
In protoplanetary disks, main sources of $g$ for small aggregates 
are stellar gravity towards the midplane of the disk and turbulent motion of the ambient gas
(see Section~3).

The charging mechanism considered in this study
 is the capturing of ionized gas particles in a weakly ionized gas \citepalias{O09}.
An important parameter characterizing the gas ionization state is the  ionization rate $\zeta$,
which is the probability that a molecule is ionized into an ion-electron pair per unit time.

\subsection{The Kinetic and Electrostatic Energies}
Collision of charged aggregates depends on 
the kinetic energy of their relative motion and the electrostatic energy.
The kinetic energy is the sum of the thermal energy ($\sim \kB T$,
where $\kB$ is the Boltzmann constant and $T$ is the temperature) and 
the energy associated with the differential drift, $E_D = [M_1M_2/(M_1+M_2)] (\Delta u_D)^2/2$,
where $M_1$ and $M_2$ are the masses of colliding aggregates.
The electrostatic energy is defined as $E_E = Q_1Q_2/(a_1+a_2)$,
where $a_{1,2}$ and $Q_{1,2}$ are the radii and charges of the aggregates.
As shown in \citetalias{OTTS11a}, the collision probability is significantly suppressed
if $E_E$ is much larger than $\kB T$ and $E_D$.
Therefore, the outcome of dust evolution is determined by 
how these energies increase as aggregates grow.

As shown in \citetalias{OTTS11a}, aggregates grow 
with a relatively narrow size distribution until the electrostatic repulsion becomes significant.
For this reason, it is useful to evaluate $E_D$ and $E_E$ 
assuming that all aggregates have the same mass at every moment. 
Following  \citetalias{OTTS11a}, we will call this the ``monodisperse approximation.'' 
Within the fractal growth regime, this is equivalent to 
treat the collision products as ballistic cluster-cluster aggregates (BCCA; e.g., \citealt*{Meakin91}).
A BCCA cluster is characterized by the monomer number $N = M/m_0$
and radius $a \approx a_0 N^{1/D}$, 
where $m_0$ and $a_0$ are the mass and radius of the monomers
and $D \approx 1.9$ is the fractal dimension of BCCA clusters 
(e.g., \citealt*{Mukai+92}; \citetalias{OTS09}). 
We simply set $D = 2$ in this paper. 
Indeed, $N$-body simulations suggest that aggregates have a fractal dimension close to 2 
unless highly unequal-sized collisions dominate their growth \citep[see][]{OTS09}.
Thus, $D \sim 2$ is a good assumption for aggregates growing 
with a relatively narrow size distribution (and without collisional compaction).

The drift velocity $\Delta u_D$ is given as follows.
Focusing on early stages of dust evolution, 
we assume that the radii of dust aggregates are much 
smaller than the mean free path of gas molecules. 
Under this assumption, the stopping time $\tau_f$ of an aggregate 
can be given by Epstein's law, 
\beq
\tau_f = \frac{3M}{4\rho_gA}\sqrt{\frac{\pi m_g}{8\kB T}},
\label{eq:tauf}
\eeq
where $\rho_g$ is the gas density, $m_g$ is the mass of gas molecules,
and $A$ is the projected area of the aggregate.
Under the monodisperse approximation, the drift energy normalized by $k_{\rm B}T$
can be written as \citepalias[see Equation~(37) of][]{OTTS11a}
\beq
{\cal E}_D \equiv \frac{E_D}{\kB T} =  f_D \eps^2 \frac{N^3}{{\cal A}(N)^2},
\label{eq:ED1} 
\eeq
where ${\cal A} = A/\pi a_0^2$ is the normalized mean projected area,
$\eps$ is the ratio of the standard deviation to the mean of the mass-to-area ratio
for given $M$, and $f_D$ is the drift energy of monomers normalized by $\kB T$.
Denoting the bulk density of monomers by $\rho_0 = 3m_0/4\pi a_0^3$,
the definition of $f_D$ reads \citepalias[Equation~(28) of][]{OTTS11a}.

\beq
f_D \equiv \frac{m_0}{2 \kB T}
\left(\frac{g \rho_0 a_0}{\rho_g}\sqrt{\frac{\pi m_g}{8 \kB T}}\right)^2.
\label{eq:fD}
\eeq
We determine ${\cal A}$ according to a fitting formula for BCCA clusters by \citet{Minato+06}, 
\beq
{\cal A}(N) =   \left\{ \begin{array}{ll}
12.5N^{0.685} \exp(-2.53/N^{0.0920}), & N < 16,  \\[3pt]
0.352N + 0.566 N^{0.862}, & N \geq 16.
\end{array} \right.
\label{eq:A_BCCA}
\eeq
Note that the mass-to-area ratio $N/{\cal A}$ approaches to a constant in the limit of $N \gg 1$.
This is a general feature of fractal aggregates with $D \la 2$ \citep[see][]{MD88,MDM89}.
In \citetalias{OTTS11a}, we have calculated $\eps$ for numerically created BCCA clusters
and found that $\eps \sim 0.1$ depending on $N$ only weakly.
We will assume $\eps = 0.1$ independently of $N$.

The dust charge $Q$  is calculated in the following way.
We introduce a dimensionless, negative surface potential $\Psi  \equiv -Qe/a\kB T$.
If the gas surrounding the dust is fully ionized, $\Psi$ is given by \citep{Spitzer41,SM02}
\beq
\frac{1}{1+\Psi} - \frac{s_i}{s_e}\sqrt\frac{m_e}{m_i}\exp\Psi = 0,
\label{eq:Psi_infty}
\eeq
where $m_{i(e)}$ is the mass of ions (electrons), 
and $s_{i(e)}$ is their sticking probability onto dust surfaces. 
We write the solution to Equation~\eqref{eq:Psi_infty} as $\Psi_\infty$. 
Assuming $m_i = 24m_{\rm H}$ (corresponding to the mass of ${\rm Mg}^+$), 
$s_i = 1$, and $s_e = 0.3$, we obtain $\Psi_\infty \approx 2.81$. 
As shown in \citetalias{OTTS11a}, the dependence of 
$\Psi_\infty$ on $(s_i/s_e)\sqrt{m_e/m_i}$ is relatively weak.
For this reason, we simply assume $\Psi_\infty = 2.81$ in this paper.

In a weakly ionized gas, $\Psi$ generally depends
on the ionization rate $\zeta$ and the size distribution of dust particles.
The value of $\Psi$ under a give condition can be calculated 
from an algebraic equation \citepalias{O09}.
We have shown in \citetalias{OTTS11a} that the solution to the equation can be well fit by 
\beq
\Psi \approx \Psi_\infty \left[ 1+ \pfrac{\Theta}{\Psi_\infty}^{-0.8} \right]^{-1/0.8}.
\label{eq:Psi}
\eeq
Here, $\Theta$ is a dimensionless quantity given by \citepalias{O09,OTTS11a}
\beq
\Theta = \frac{\zeta n_ge^2}{A_{\rm tot}C_{\rm tot}\kB T}\sqrt{\frac{\pi m_i}{8 \kB T}},
\label{eq:Theta}
\eeq
where $n_g = \rho_g/m_g$ is the number density of gas molecules, 
and $A_{\rm tot}$ and $C_{\rm tot}$ is the total projected area 
and total radius of dust aggregates within a unit volume, respectively (note that we have assumed $s_i = 1$).
In general, $\Theta$ decreases with decreasing $\zeta$ and increasing the amount of dust. 
As found from Equation~\eqref{eq:Psi}, 
$\Psi$ approaches to $\Psi_\infty$ only in the limit $\Theta \gg \Psi_\infty$
and decreases with $\Theta$ in the opposite limit.
This reflects the fact that gas ionization is insufficient for dust to be fully charged 
when $\Theta$ is small \citepalias{O09}.
Following \citetalias{O09}, we will refer to the gas--dust mixture
as the ion--electron plasma (IEP) and the ion--dust plasma (IDP)
 when $\Theta \ga \Psi_\infty$ and $\Theta \la \Psi_\infty$, respectively.

If dust aggregates are monodisperse, we can write $A_{\rm tot} = An_0/N$
and $C_{\rm tot} = an_0/N$, where $n_0$ is the number density of dust monomers
(note that $n_0/N$ represents the number density of aggregates).
Using these expressions as well as the scaling
${\cal A} = A/\pi a_0^2$ and $a = a_0N^{1/D} = a_0N^{1/2}$ described above, 
Equation~\eqref{eq:Theta} reduces to
\beq
\Theta = \frac{ h N^{3/2}}{{\cal A}(N)} \Psi_\infty,
\label{eq:Theta1}
\eeq
where 
\beq
h \equiv \frac{\zeta n_g e^2}{\pi a_0^3 n_{0}^2 \Psi_\infty \kB T}\sqrt\frac{\pi m_i}{8\kB T}
\label{eq:h}
\eeq
is a dimensionless quantity representing the ionization rate of the gas.

The electrostatic energy for monodisperse aggregates 
is given by \citepalias[see Equations~(27) and (29) of][]{OTTS11a}
\beq
{\cal E}_E \equiv \frac{E_E}{\kB T} 
= \frac{f_E}{2}\pfrac{\Psi}{\Psi_\infty}^2N^{1/2},
\label{eq:EE} 
\eeq
where 
\beq
f_E \equiv \frac{\Psi_\infty^2 a_0 \kB T}{e^2}
\label{eq:fE}
\eeq
is the electrostatic energy between monomers in the IEP state  (normalized by $\kB T$).
If we use Equations~\eqref{eq:Psi} and \eqref{eq:Theta1}, 
Equation~\eqref{eq:EE} can be rewritten as 
\beq
{\cal E}_E = \frac{f_E}{2}\left[ 1+ \pfrac{h N^{3/2}}{{\cal A}(N)}^{-0.8} \right]^{-2.5}N^{1/2}.
\label{eq:EE1} 
\eeq

Thus, the normalized energies ${\cal E}_D$ and ${\cal E}_E$ 
are characterized by three dimensionless parameters ($f_D$, $f_E$, $h$).

\subsection{The Drift Mass and Plasma Transition Mass}
It is useful to introduce two critical masses 
characterizing the motion and charge state of aggregates.
The first one is the drift mass $M_D(\equiv N_Dm_0)$ 
defined by ${\cal E}_D(N_D) = 1$ (or equivalently, $E_D(M_D) = \kB T$).
This represents the mass at which the differential drift 
begins to take over Brownian motion in the collision velocity.
Using Equation~\eqref{eq:ED1}, the equation for $N_D$ can be rewritten as
\beq
\frac{{\cal A}(N_D)^2}{N_D^3} = f_D \eps^2,
\label{eq:ND_def}
\eeq
Thus, $N_D$ is a function of $f_D\eps^2$.
As shown in~\citetalias{OTTS11a}, $N_D$  increases with decreasing $f_D\eps^2$
and behaves as $N_D \approx 1/b^2f_D\eps^2$ in the limit of $f_D\eps^2 \ll 1$,
where $b = 1/0.352 \approx 2.84$ is the asymptotic value of $N/{\cal A}$ in the limit 
of $N \gg 1$ (see Equation~\eqref{eq:A_BCCA} and the remark below the equation).
Note that $N_D$ approaches unity as $f_D\eps^2 \to 1$
and becomes ill-defined at higher $f_D\eps^2$.
For this reason, we simply set $N_D = 1$ when $f_D\eps^2>1$.

The second critical mass is the plasma transition mass $M_P(\equiv N_Pm_0)$
defined by $\Theta(N_P) = \Psi_\infty$. 
Since $\Theta$  increases  with $N$, 
the gas--dust mixture is in the IDP state when $N \ll N_P$ and is in the IEP state when $N \gg N_P$.
With Equation~\eqref{eq:Theta1}, the equation for $N_P$ can be rewritten as 
\beq
\frac{{\cal A}(N_P)}{N_P^{3/2}} = h.
\label{eq:NP_def}
\eeq
The plasma transition mass increases with decreasing $h$
and behaves as $N_P \approx 1/b^2h^2$ in the limit of $h \ll 1$.
Hence, if $N$ and $N_P \gg 1$, we can approximate Equation~\eqref{eq:EE1} as  
\beqn
{\cal E}_E &\approx&
 \frac{f_E}{2}\left[ 1+ \pfrac{N^{1/2}}{N_P}^{-0.8} \right]^{-2.5}N^{1/2} 
 \nonumber \\
&\approx&  \left\{ \begin{array}{ll}
\dfrac{f_E}{2} \dfrac{N^{3/2}}{N_P},  & N \ll N_P, \\
\dfrac{f_E}{2}N^{1/2},  & N \gg N_P. 
\end{array} \right.
\label{eq:EE_asympt}
\eeqn
We will use this approximation in the following subsection.
Again, we set $N_P = 1$ when $h > 1$.

\subsection{The Growth Criteria}
Now we are ready to write down the growth criteria. 
The first criterion is given by \citepalias[Equation~(57) of][]{OTTS11a}
\beq
{\cal E}_E(N_D) \ga 6.
\label{eq:growthcond}
\eeq
If this inequality holds, all dust aggregates stop growing at a certain size (``total freezeout'').
The size of the ``frozen'' aggregates is characterized by
the freezeout mass $M_F(\equiv N_Fm_0)$ defined by ${\cal E}_E(N_F) = {\cal E}_K(N_F)$,
where ${\cal E}_K \equiv 1+ {\cal E}_D$ is the total kinetic energy (i.e., thermal energy + drift energy).
As shown in \citetalias{OTTS11a}, $M_F$ is smaller than $M_D$
whenever Equation~\eqref{eq:growthcond} holds  
(i.e., the total freezeout occurs only in the Brownian motion regime),
so the definition of $M_F$ is effectively equivalent to ${\cal E}_E(N_F) \approx 1$.

When the inequality in Equation~\eqref{eq:growthcond} is reversed,
the outcome of dust growth depends on the second criterion
\citepalias[Equation~(58) of][]{OTTS11a}
\beq
\Psi_\star \equiv \frac{2\Psi_\infty}{(f_E^2 N_D)^{1/4}} \la \frac{\Psi_\infty}{4}.
\label{eq:Psi_star}
\eeq
If this inequality holds, a certain number of aggregates continue growing
while the rest stop growing at $M \approx M_D$ (``bimodal growth'').
Otherwise, all aggregates continue growing
with a relatively narrow size distribution (``unimodal growth'').
The growth criteria are summarized in Table~\ref{table1}.
\begin{deluxetable}{ccl}
\tablecaption{Three Outcomes of the Growth of Charged Dust}
\tablecolumns{3}
\tablehead{
\multicolumn{2}{c}{Conditions} & \colhead{Outcome}
}
\startdata
${\cal E}_E(N_D) \ga 6$ & \nodata & Total freezeout \\
${\cal E}_E(N_D)  \la 6 $ & $\Psi_\star \la \Psi_\infty/4$ &  Bimodal growth \\
${\cal E}_E(N_D)  \la 6 $ & $\Psi_\star \ga \Psi_\infty/4$ & Unimodal growth
\enddata
\label{table1}
\end{deluxetable}

For later convenience, we derive an approximate expression of ${\cal E}_E(N_D)$ 
applicable for $f_D\eps^2, h \ll 1$.
Recall that  $N_D (\approx 1/b^2f_D\eps^2)$ 
and $N_P (\approx 1/b^2h^2)$ are much larger than unity if $f_D\eps^2 \ll 1$
and $h\ll 1$, respectively (see Section 2.2).
Therefore, we can approximate ${\cal E}_E(N_D)$ by Equation~\eqref{eq:EE_asympt}.
Thus, we obtain 
\beq
{\cal E}_E(N_D) \approx  \left\{ \begin{array}{ll}
\dfrac{f_E}{2}N_D^{1/2} \approx \dfrac{f_E}{2b \eps f_D^{1/2}}, & N_D \gg N_P, \\
\dfrac{f_E}{2} \dfrac{N_D^{3/2}}{N_P} \approx \dfrac{f_E h^2}{2b \eps^3 f_D^{3/2}}, & N_D \ll N_P, \\\end{array} \right.
\label{eq:EEND_asympt}
\eeq
for $f_D\eps^2$, $h \ll 1$.
Equation~\eqref{eq:EEND_asympt} is useful because 
${\cal E}_E(N_D)$ is explicitly given as a function of ($f_D\eps^2$, $f_E$, $h$).

\section{Disk Model}
The criteria shown in the previous section enables us 
to examine how small dust aggregates evolve at each location in a protoplanetary disk.
Here, we introduce a disk model used in this paper.
The model is essentially the same as the one adopted in \citetalias{O09}.

\subsection{Structure of the Gas Disk}
We assume the gas surface density $\Sigma_g$ obeying a power law
\beq
\Sigma_g(r) = 1.7\times 10^3\eta_\Sigma\pfrac{r}{1~\AU}^{-3/2}  {\rm~g~cm^{-2}},
\label{eq:Sigma_g}
\eeq
where $r$ is the distance from the central star,
and $\eta_\Sigma$ is a scaling factor.
The model with $\eta_\Sigma = 1$
is known as the minimum-mass solar nebula (MMSN) model of \citet{Hayashi81}.
We leave $\eta_\Sigma$ as a free parameter to clarify the dependence on the disk mass.

The gas temperature $T$ is assumed to be isothermal in the vertical direction,
and the radial profile of $T$ is given by that of \citet{Hayashi81},
\beq
T = 280\pfrac{r}{1~\AU}^{-1/2} {~\rm K}.
\label{eq:T}
\eeq
In reality, the temperature can be lower because of the large optical thickness provided by dust
and can be higher because of the turbulent heating.
However, we ignore these effects for simplicity.

To obtain the vertical structure of the disk,
we assume hydrostatic equilibrium of the gas in the vertical direction.
This leads to 
\beq
\rho_g = \frac{\Sigma_g}{\sqrt{2\pi}H}\exp\left(-\frac{z^2}{2H^2}\right),
\label{eq:rhog}
\eeq
where $z$ is the height from the disk midplane and 
\beq
H \equiv \frac{c_s}{\Omega_{\rm K}}
\label{eq:H}
\eeq
is the gas scale height.
The isothermal sound velocity $c_s$ and the Keplerian orbital frequency $\Omega_{\rm K}$
are given by 
\beq
c_s = \sqrt{\frac{\kB T}{m_g}}
\label{eq:cs}
\eeq
 and 
\beq
\Omega_{\rm K} = \sqrt{\frac{GM_*}{r^3}},
\label{eq:Omega}
\eeq
where $G$ is the gravitational constant and $M_*$ is the mass of the central star.
We assume a mean molecular weight of $2.34$ and write $m_g = 2.34m_{\rm H}$,
where $m_{\rm H}$ is the hydrogen mass. 
For the stellar mass, we assume $M_*=1M_\sun$. 

Dust material is assume to be well mixed in the disk,
and the dust density $\rho_d$ is related to $\rho_g$ as 
\beq
\rho_d(r,z) = f_{dg}\rho_g(r,z),
\label{eq:rhod}
\eeq
where $f_{dg}$ is the dust-to-gas mass ratio.
We choose $f_{dg} = 0.014$ as calculated from
the solar system abundance of condensates including water ice \citep{Pollack+94}.
The bulk density $\rho_0$ of dust monomers is set to $1.4{\rm~g~cm^{-3}}$ 
consistently with the adopted solar system abundance.
We ignore the sublimation of water ice in inner disk regions for simplicity.

\subsection{Dust Motion}
We consider vertical sedimentation and disk turbulence
as the mechanism driving differential drift between aggregates.
Assuming  $z\ll r$, the vertical component of the stellar gravity is given by 
\beq
g_S = \Omega_{\rm K}^2 z.
\label{eq:gS}
\eeq
The relative velocity driven by turbulence generally depends on 
the ratio of the stopping times ($\tau_{f,1}$ and $\tau_{f,2}$) of colliding aggregates 
to the turnover times of turbulent eddies \citep{OC07}.
In particular, if the stopping times are shorter than the turnover time $t_\eta$ 
of the smallest eddies, 
the turbulence-driven relative speed can be approximately 
written in the form $g_T|\tau_{f,1}-\tau_{f,2}|$,
where
\beq
g_T \approx u_\eta/t_\eta
\label{eq:gT_def}
\eeq
is the effective acceleration driven by turbulence,
and $u_\eta$ is the characteristic velocity of the smallest eddies \citep{W84,OC07}.
In this case, the total drift acceleration $g$ is given by 
\beq
g^2 = g_S^2 + g_T^2,
\eeq
where $g_S$ and $g_T$ are the vertical gravity and 
the effective acceleration driven by turbulence, respectively.
The validity of the strong-coupling approximation (i.e., $\tau_{f,1}$, $\tau_{f,2} \ll t_\eta$)
is discussed later.

We evaluate $g_T$ in the following way.
We express the strength of turbulence with the turbulent viscosity 
$\nu_{\rm turb} = \alpha c_s^2/\Omega_{\rm K}$,
where $\alpha$ is the so-called alpha parameter for turbulence.
The turbulent viscosity can be alternatively written as $\nu_{\rm turb} = u_L^2 t_L$,
where $u_L$ and $t_L$ represent the characteristic velocity and correlation time 
of the largest eddies, respectively. 
We assume $t_L = 1/\Omega_{\rm K}$ as is for magnetorotational  turbulence 
\citep[e.g.,][]{FP06,TWBY06}.
Equating the two expressions for $\nu_{\rm turb}$, 
the characteristic velocity is obtained as $u_L = \sqrt{\alpha}c_s$.
Assuming the Kolmogorov spectrum, $u_\eta$ and $t_\eta$ are given 
in terms of $u_{ L}$ and $t_{L}$
as $u_\eta = {\rm Re}^{-1/4}u_{L}$ and $t_\eta = {\rm Re}^{-1/2}t_{L}$,
where  ${\rm Re} = \nu_{\rm turb}/\nu_{\rm mol}$ is the Reynolds number.
The molecular viscosity $\nu_{\rm mol}$ is given by 
$\nu_{\rm mol} = u_g/(2n_g\sigma_{\rm mol})$,
where $u_g = \sqrt{8/\pi}c_s$ is the molecular thermal speed
 and $\sigma_{\rm mol} = 2\times 10^{-15}~{\rm cm^2}$ 
 is the molecular collision cross section \citep{CC70}.
Using these relations, we can rewrite the turbulence-driven acceleration $g_T$
(Equation~\eqref{eq:gT_def}) as
\beq
g_T \approx \sqrt{\alpha}{\rm Re}^{1/4}c_s\Omega_{\rm K}.
\label{eq:gT}
\eeq 
Note that $g_T \propto \alpha^{3/4}$ because ${\rm Re} \propto \alpha$.
In the following section, we treat $\alpha$ as a free parameter.

Now let us check the validity of the strong-coupling approximation
for the turbulence-driven relative velocity.
It is useful to rewrite the Reynolds number as 
\beq
{\rm Re} = \frac{\alpha \Sigma_g \sigma_{\rm mol}}{2m_g}{\rm e}^{-z^2/2H^2},
\label{eq:Re}
\eeq 
where we have used $\nu_{\rm turb} = \alpha c_s H$, 
$\nu_{\rm mol} =\sqrt{2/\pi}c_sm_g/(\rho_g\sigma_{\rm mol})$, and 
Equation~\eqref{eq:rhog}. 
Substituting this expression and Equation~\eqref{eq:Sigma_g}
into $t_\eta = {\rm Re}^{-1/2}t_\eta = {\rm Re}^{-1/2}/\Omega_{\rm K}$, we have
\beq
\Omega_{\rm K}t_\eta 
\approx 2\times 10^{-5}f_\Sigma^{-1/2}\pfrac{\alpha}{10^{-2}}^{-1/2}\pfrac{r}{1~\AU}^{3/4}{\rm e}^{z^2/4H^2}.
\label{eq:Omegat_eta}
\eeq
Note that ${\rm Re}$ (and hence $\Omega_{\rm K}t_\eta$)
is independent of $T$ and $\Omega_{\rm K}$.
For the stopping time $\tau_f$, we use the fact that
the mass-to-area ratio of fractal $(D\sim 2)$ aggregates 
approaches to a constant in the limit $N \gg 1$ (see Section 2.2).
Substituting Equations~\eqref{eq:rhog}--\eqref{eq:cs} into Equation~\eqref{eq:tauf}
and using $N/{\cal A} \la b \approx 2.84$, 
we obtain 
\beqn
\Omega_{\rm K}\tau_f &\la& \frac{b \pi}{2}\frac{\rho_0a_0}{\Sigma_g}{\rm e}^{z^2/2H^2}
\nonumber \\
&\approx& 4\times 10^{-8}f_\Sigma^{-1}\pfrac{a_0}{0.1~\micron}\pfrac{r}{1~\AU}^{3/2}
{\rm e}^{z^2/2H^2}.
\label{eq:Omegatau_f}
\eeqn
Comparing Equation~\eqref{eq:Omegatau_f} with Equation~\eqref{eq:Omegat_eta},  
we find that the strong coupling approximation is valid (i.e., $\tau_f\ll t_\eta$)
as long as $\alpha \la 10^{-2}$, $a_0\la 1~\micron$, and $r\la 100~\AU$.

\subsection{Gas Ionization}
For ionizing sources, we consider Galactic cosmic rays \citep{UN81}, 
stellar X-rays \citep{IG99}, and radionuclides \citep{UN09}.
Thus, we decompose the ionization rate as 
$\zeta \approx \zeta_{\rm CR} + \zeta_{\rm XR} + \zeta_{\rm RA}$
where $\zeta_{\rm CR}$, $\zeta_{\rm XR}$, and $\zeta_{\rm RA}$ denote the rate of ionization by 
cosmic rays, X-rays, and radionuclides, respectively.
We do not consider thermal ionization because it is negligible
at  gas temperatures $\ll 10^3~{\rm K}$, 
or at heliocentric distances $\gg 0.1~\AU$ \citep{Umebayashi83}.

The cosmic-ray ionization rate is given by a fitting formula \citep{UN09}
\beqn
\zeta_{\rm CR}(r,z) &=&  \frac{\zeta_{\rm CR,0}}{2}
\left\{ {\rm e}^{-\Sigma_g^+(r,z)/\Sigma_{\rm CR}}
\left[1+\pfrac{\Sigma_g^+(r,z)}{\Sigma_{\rm CR}}^\frac{3}{4}\right]^{-\frac{4}{3}} \right. \nonumber \\
&&+ \left. {\rm e}^{-[\Sigma_g(r)-\Sigma_g^+(r,z)]/\Sigma_{\rm CR}}
\left[1+\pfrac{\Sigma_g(r)-\Sigma_g^+(r,z)}{\Sigma_{\rm CR}}^\frac{3}{4}\right]^{-\frac{4}{3}} \right\}, \nonumber \\
\label{eq:zetaCR}
\eeqn
where $\zeta_{\rm CR,0} = 1.0\times 10^{-17}{\rm~s^{-1}}$ 
is the cosmic-ray ionization rate in the interstellar space, 
$\Sigma_{\rm CR} \approx 96{\rm~g~cm^{-2}}$ is the attenuation length of the ionization rate,
and 
\beq
\Sigma_g^+(r,z) \equiv \int_z^\infty\rho_g(r,z')dz' 
= \frac{\Sigma_g}{2} {\rm erfc}\pfrac{z}{\sqrt{2} H}
\label{eq:Sigma+}
\eeq
is the vertical gas column density above altitude $z$.

For the radionuclide ionization, we assume 
$\zeta_{\rm RA} \approx 7 \times 10^{-19}{\rm~s^{-1}}$, 
which corresponds to the ionization rate by
a short-lived radionuclide ${\rm ^{26}Al}$ \citep{UN09}
with the abundance ratio of ${\rm ^{26}Al/^{27}Al} = 5 \times 10^{-5}$ \citep{Lee+77}.
We neglect other short-lived radionuclides and all long-lived ones
since they give only minor contributions.
We also neglect the decrease in ${\rm ^{26}Al}$ due to the radioactive decay;
this can be done as long as we consider an early stage of dust evolution 
within a timescale of $10^6~{\rm yr}$.

The stellar X-ray ionization rate has been calculated by \citet{IG99} 
using the Monte Carlo radiative transfer code including Compton scattering. 
A useful fitting formula is given by \citet{TS08},
\beqn
\zeta_{\rm XR}(r,z) &\approx& \zeta_{\rm XR,0}
\pfrac{r}{1~\AU}^{-2}\pfrac{L_{\rm XR}}{2\times 10^{30}{\rm~erg~s^{-1}}} \nonumber \\
&&\times
\left( {\rm e}^{-\Sigma_g^+(r,z)/\Sigma_{\rm XR}}
+ {\rm e}^{-[\Sigma_g(r)-\Sigma_g^+(r,z)]/\Sigma_{\rm XR}} \right), \nonumber \\ 
\label{eq:zetaXR}
\eeqn
where $L_{\rm XR}$ is the X-ray luminosity of the central star, and 
$\zeta_{\rm XR,0} = 2.6\times 10^{-15}{\rm~s^{-1}}$ and $\Sigma_{\rm XR} = 8.0 {\rm~g~cm^{-2}}$ 
are the fitting parameters.
This fitting formula approximately reproduces the $\kB T_{\rm XR} = 5~{\rm keV}$ result of \citet{IG99}
at $\Sigma_g^+ \ga 1{\rm~g~cm^{-2}}$
where scattered hard ($\ga 5~{\rm keV}$) X-rays are responsible for the ionization.
At higher altitudes, Equation~\eqref{eq:zetaXR} underestimates the ionization rate
since it ignores the contribution of softer X-rays.
We nevertheless use Equation~\eqref{eq:zetaXR} in this paper 
because the critical energy ${\cal E}_E(N_D)$ is independent of $\zeta$ at 
such high altitudes (see Equation~\eqref{eq:EEND_approx}).
We take $L_{\rm XR} = 2\times 10^{30}{\rm~erg~s^{-1}}$ in accordance with the median characteristic
X-ray luminosity observed by {\it Chandra} for young solar-mass stars in the Orion Nebula Cluster \citep{Wolk+05}.
Although the characteristic X-ray temperature 
$\kB T_{\rm XR} \approx 2.4~{\rm keV}$ observed by \citet{Wolk+05} is lower than 
the assumed value of $5~{\rm keV}$, 
the choice of the temperature does not significantly affect the resulting ionization rate \citep{IG99}. 

\begin{figure}
\plotone{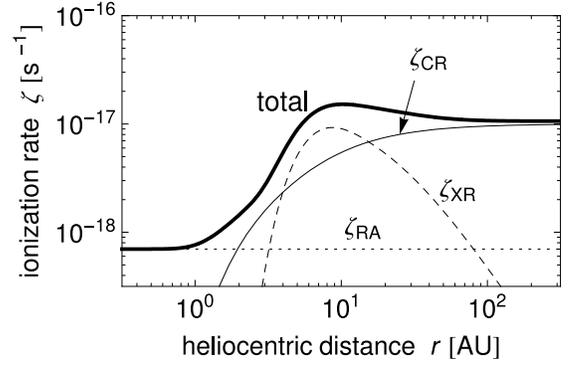}
\caption{
Radial profile of the ionization rate at $z = H$ for  $\eta_\Sigma = 1$.
The thick solid curve shows the total ionization rate $\zeta$,
while the thin solid, dashed, and dotted curves show the contribution
 from Galactic cosmic rays ($\zeta_{\rm CR}$),
stellar X-rays ($\zeta_{\rm XR}$), and radionuclides ($\zeta_{\rm RA}$).
}
\label{fig:zeta}
\end{figure}
As an example, Figure~\ref{fig:zeta} shows
the radial profile of the ionization rate measured at $z=H$ for $\eta_\Sigma = 1$. 
Cosmic rays dominate ionization at $r \ga 20~\AU$,
while X-rays dominate at $4~\AU \la r \la 20~\AU$.
At $r \la 1~\AU$,  both cosmic rays and X-rays are significantly attenuated, 
and thus the ionization rate reaches the floor value $\zeta_{\rm RA}$.

\subsection{Global Profiles of $f_D$, $f_E$, and $h$}
With the disk model described above, 
we can obtain analytical expressions for $f_D$, $f_E$, and $h$ as a function of $r$ and $z$.

First, we decompose $f_D$ (Equation~\eqref{eq:fD}) into two components,
\beq
f_D = f_{D,S} + f_{D,T},
\label{eq:fDr}
\eeq
where $f_{D,S}$ and $f_{D,T}$ are the contributions from stellar gravity an turbulence
(i.e., $f_D$ with $g = g_S$ and $g=g_T$), respectively. 
Using Equations~\eqref{eq:rhog}--\eqref{eq:cs} and \eqref{eq:gS}, we can rewrite $f_{D,S}$ as
\beqn
f_{D,S} &=& \frac{9\pi^2}{128}\frac{m_0}{m_g}\pfrac{m_0}{\pi a_0^2 \Sigma_g}^2
\pfrac{z}{H}^2 {\rm e}^{z^2/H^2}\nonumber \\
&\approx& 4\times 10^{-7} \pfrac{a_0}{0.1~\micron}^5 \pfrac{\Sigma_g}{10^3{\rm~g~cm^{-2}}}^{-2}
\pfrac{z}{H}^2{\rm e}^{z^2/H^2},  \quad
\label{eq:fDSr}
\eeqn
where the radial profile of $\Sigma_g$ is given by Equation~\eqref{eq:Sigma_g}.
For $f_{D,T}$ we use Equations~\eqref{eq:rhog}--\eqref{eq:cs}, \eqref{eq:gT}, and \eqref{eq:Re}
to obtain
\beqn
f_{D,T} &=& \frac{9\pi^2}{128}\frac{m_0}{m_g}\alpha\sqrt{\rm Re}\pfrac{m_0}{\pi a_0^2 \Sigma_g}^2
 {\rm e}^{z^2/H^2}\nonumber \\
&\approx& 2 \times 10^{-4} \pfrac{\alpha}{10^{-2}}^{3/2}
\pfrac{a_0}{0.1~\micron}^5 \pfrac{\Sigma_g}{10^3{\rm~g~cm^{-2}}}^{-3/2}  {\rm e}^{3z^2/4H^2}.
\nonumber \\
\label{eq:fDTr}
\eeqn
Note that both $f_{D,S}$ and $f_{D,T}$ are independent of $T$ and $\Omega_{\rm K}$ for fixed $z/H$.

For $f_E$, Equation~\eqref{eq:fE} directly gives
\beq
f_E \approx 14\pfrac{a_0}{0.1~\micron}\pfrac{T}{300~{\rm K}},
\label{eq:fEr}
\eeq
where the temperature profile is given by Equation~\eqref{eq:T}.
Finally, for $h$, substitution of Equations~\eqref{eq:rhog}--\eqref{eq:rhod}
and $n_0 = \rho_d/m_0 $ into Equation~\eqref{eq:h} leads to
\beqn
h &=& \frac{\pi}{2}f_{dg}^{-2} \sqrt{\frac{m_i}{m_g}}\frac{m_0}{m_g}
\frac{m_0}{\pi a_0^2\Sigma_g}\frac{e^2}{\Psi_\infty a_0 \kB T}\frac{\zeta}{\Omega_{\rm K}} 
{\rm e}^{z^2/2H^2}
\nonumber \\
&\approx& 7\times 10^{-6} \pfrac{a_0}{0.1~\micron}^3\pfrac{T}{300~{\rm K}}^{-1}\pfrac{\Sigma_g}{10^3{\rm~g~cm^{-2}}}^{-1}
 \nonumber \\
&&\times 
\pfrac{\zeta}{10^{-17}~{\rm s^{-1}}} \pfrac{2\pi/\Omega_{\rm K}}{1~{\rm yr}}{\rm e}^{z^2/2H^2}.
\label{eq:hr}
\eeqn

\begin{figure}
\plotone{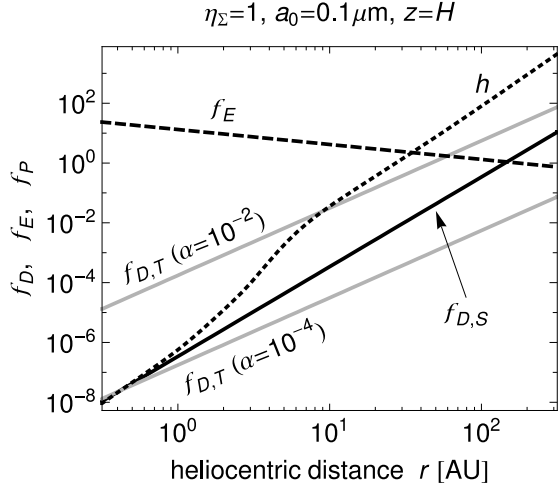}
\caption{
Radial profiles of $f_{D,S}$ (Equation~\eqref{eq:fDSr}; solid black line), 
$f_E$ (Equation~\eqref{eq:fEr}; dashed line), and $h$ (Equation~\eqref{eq:hr}; dotted line) 
at $z=H$ for $\eta_\Sigma = 1$ and $a_0 = 0.1~\micron$.
The two gray lines show $f_{D,T}$ (Equation~\eqref{eq:fDTr})
for $\alpha = 10^{-4}$ and $10^{-2}$. 
}
\label{fig:param}
\end{figure}
As an example, we plot in Figure~\ref{fig:param} the radial profiles of 
$f_{D,S}$, $f_{D,T}$, $f_E$, and $h$ at $z=H$ for $\eta_\Sigma =1$ and $a_0 = 0.1~\micron$.
It is seen that the ratio of $f_{D,T}/f_{D,S}$ decreases with $r$,
meaning that the effect of turbulent is relatively insignificant at outer parts of the disk.
This is because the ratio $f_{D,T}/f_{D,S}$ is proportional to ${\rm Re}^{1/2}$
and the Reynolds number ${\rm Re} \propto \Sigma_{\rm g}$
decreases with $r$ for fixed $\alpha$ and $z/H$.

\section{Results}
\subsection{Fiducial Case}
\begin{figure}
\epsscale{1.15}
\plotone{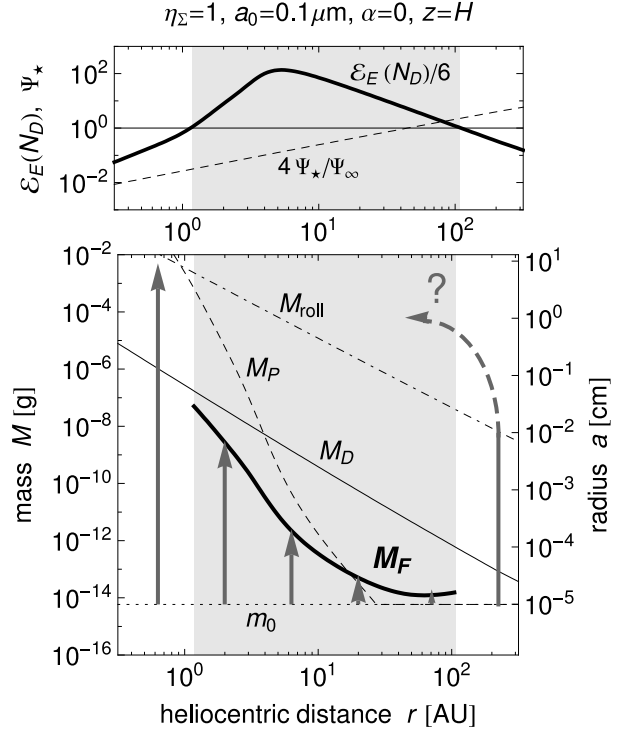}
\vspace{-0.5cm}
\caption{
Upper panel: radial profile of 
${\cal E}_E(N_D)$ (solid curve) and $\Psi_\star$ (dashed curve)
at $z=H$ for the fiducial model.
At heliocentric distances where ${\cal E}_E(N_D) \ga 6$ (shaded area),
fractal dust growth stalls at mass $M \approx M_F$ because of the electrostatic barrier.
Lower panel: freezeout mass $M_F$ (thick solid line), drift mass $M_D$ (thin solid line),
plasma transition mass $M_P$ (dashed line), and critical restructuring mass $M_{\rm roll}$ 
(dot-dashed line) at $z=H$ for the same model.
The initial (monomer) mass $m_0$ is shown by the dotted line.  
The gray arrows illustrate the growth history of dust growth starting from various locations.
Aggregates grown beyond $M=M_{\rm roll}$  leave the fractal growth regime
and will finally drift inwards as schematically shown by the dashed curved arrow (see also Section 5.2.2).
}
\label{fig:EN0}
\end{figure}

As the fiducial example, we begin with the case of $\eta_\Sigma = 1$ (i.e., the original MMSN),
$\alpha=0$ (i.e., laminar disk), and $a_0 = 0.1~\micron$.

The upper panel of Figure~\ref{fig:EN0} shows the radial profiles 
of  ${\cal E}_E(N_D)$ and $\Psi_\star$ calculated at one scale height above the midplane ($z= H$).
We find that ${\cal E}_E(N_D)$ exceeds the critical value 6 at $1~\AU \la r \la 100~\AU$.
The total freezeout occurs in this wide region.
The outcome of dust growth outside the frozen zone depends on the value of $\Psi_\star$ 
(Equation~\eqref{eq:Psi_star}).
As seen in the figure, $\Psi_\star$  increases  with $r$ 
and exceeds the critical value $\Psi_\infty/4$ at $r \approx 50~\AU$. 
Hence, the growth is unimodal at $r\ga 100~\AU$ and is bimodal at $r \la~1\AU$.

The lower panel of Figure~\ref{fig:EN0} plots the freezeout mass $M_F$ in the frozen zone
as well as the drift mass $M_D$ and the plasma transition mass $M_P$ at $z=H$.
It is seen that $M_F$ rapidly decreases as $r$ increases,
with the maximum value $M_F \sim 10^{-7}~{\rm g}$ ($a \sim 0.3~{\rm mm}$) 
and the minimum value as low as a few monomer masses.

To see how early dust evolution stalls in the frozen zone, 
we compare $M_F$ with the threshold mass $M_{\rm roll}$ for the onset of collisional compaction.
The threshold mass is defined by $E_K (M_{\rm roll}) = E_{\rm roll}$, where 
\beqn
E_{\rm roll} &=& 3\pi^2 \gamma a_0\xi_{\rm crit} \nonumber \\
&\approx& 6 \times 10^{-10} \pfrac{\gamma}{100~{\rm erg~cm^{-2}}}
\pfrac{\xi_{\rm crit}}{2~{\rm\mathring{A}}}\pfrac{a_0}{0.1~\micron}~{\rm erg}\qquad
\label{eq:Eroll}
\eeqn
is the energy needed to roll one monomer on another in contact by $90^\circ$,
$\gamma$ is the surface adhesion energy for the two monomers,
and $\xi_{\rm crit}$ is the critical tangential displacement for starting the rolling \citep{DT95,DT97}.
Numerical simulations \citep{DT97,SWT08} and laboratory experiments \citep{BW00}
show that dust grows into fractal aggregates as long as
 the collision energy $E_K$ is below $E_{\rm roll}$.
Therefore, we can regard $M_{\rm roll}$ as the maximum mass below which 
the collisional compaction can be neglected.
For icy monomers, $\gamma$ is estimated as $100 {\rm~erg~cm^{-2}}$ \citep{I92} 
but a realistic value of $\xi_{\rm crit}$ is unknown.
For a conservative estimation, we assume the minimum displacement 
$\xi_{\rm crit} = 2~{\rm\mathring{A}}$ anticipated by the microscopic theory \citep{DT97}, 
which makes our aggregates the most easily compressed.
In the lower panel of Figure~\ref{fig:EN0}, we plot $M_{\rm roll}$ as a function of $r$.
We see that $M_F$ is at least four orders of magnitude smaller than $M_{\rm roll}$.
This is a robust result because Brownian motion dominates the relative motion of frozen aggregates
and because the thermal energy $\sim \kB T$ is generally much smaller than $E_{\rm roll}$.
Note that the electrostatic barrier is in marked contrast to the other known growth barriers 
(e.g., radial drift, fragmentation, bouncing) which obstruct dust growth 
{\it after} the collisional compaction becomes effective.
  
It is worth mentioning that the frozen zone always has inner and outer boundaries at finite $r$. 
In outer regions where $M_D \gg M_P$ (see Figure~\ref{fig:EN0}),
the approximate formula for ${\cal E}_E(N_D)$ (Equation~\eqref{eq:EEND_asympt}) reads
\beqn
{\cal E}_E(N_D)
&\approx&  \frac{f_E}{2b \eps f_D^{1/2}}  \nonumber \\
&\approx& 8\pfrac{a_0}{0.1~\micron}^{-3/2}\pfrac{\Sigma_g}{2~{\rm g~cm^{-2}}}
\pfrac{T}{30{\rm~K}} \frac{H}{z}{\rm e}^{-z^2/2H^2}, \nonumber \\
\label{eq:EEND_out}
\eeqn
where we have used Equations~\eqref{eq:fDSr} and \eqref{eq:fEr}.
Note that ${\cal E}_E(N_D)$ at fixed  $z/H$ decreases as $r$ {\it increases}
because $\Sigma_g $ and $T$ generally decrease with increasing $r$.
On the other hand, in inner regions where $M_D \ll M_P$, 
 ${\cal E}_E(N_D)$ is smaller than that in Equation~\eqref{eq:EEND_out} 
by a factor $M_D/M_P \propto \zeta^2/\Omega_{\rm K}^{2}T^{2} \propto \zeta^2 r^3/T^{2}$
 (see Equation~\eqref{eq:EEND_asympt}). 
In this region,  ${\cal E}_E(N_D) \propto \Sigma_g\zeta^2r^3/T$
decreases as $r$ {\it decreases} (unless $\Sigma_g/T$ is steeper than $r^{-3}$), 
since the ionization rate $\zeta$ generally  decreases as the surface density increases.
From these facts, we find that the frozen zone always has inner and outer boundaries
at finite $r$.

\begin{figure}
\plotone{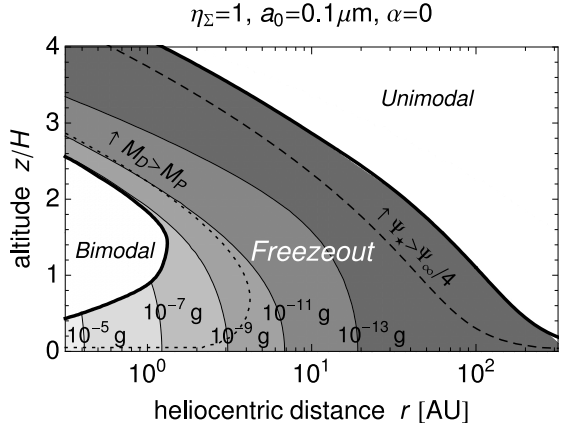}
\caption{
Two-dimensional map showing the location of the frozen zone in the fiducial disk model.
Shaded is the ``frozen'' zone, where the total freezeout occurs (Equation~\eqref{eq:growthcond}).
In this region, fractal dust growth stalls at the freezeout mass $M_F$
(whose value is shown by the contours).
The second growth criterion (Equation~\eqref{eq:Psi_star}) is satisfied above the dashed curve,
indicating that the unimodal growth occurs outside above the frozen zone
while the bimodal growth occurs inside the frozen zone.
Outside the dotted curve, the drift mass $M_D$ is larger than the plasma transition mass $M_P$,
so the approximate formula for ${\cal E}_E(N_D)$ (Equation~\eqref{eq:EEND_out}) is applicable.
}
\label{fig:rz}
\end{figure}
The two-dimensional ($r$--$z$) map of the frozen zone is displayed in Figure~\ref{fig:rz}.
At $z \ga H$, the frozen zone moves inwards as $z$ increases,
because  ${\cal E}_E(N_D)$ is lower at larger $z$ and $r$ (see Equation~\eqref{eq:EEND_out}). 
A region very close to the midplane is entirely covered by the frozen zone 
because the settling velocity vanishes there (i.e., ${\cal E}_D \to 0$).
The bimodal growth zone covers middle altitudes at $r \la 1~\AU$ and expands towards smaller $r$.
The upper boundary between the bimodal and frozen zones is roughly characterized 
by the column depth $\Sigma_g^+(z) \approx 10^{2}~{\rm g~cm^{-2}}$.
This reflects the fact that cosmic rays is significantly attenuated at such depths.
The unimodal growth is only allowed at large $r$ and $z$ where the settling velocity 
is high enough for dust to overcome the electrostatic barrier.

\begin{figure}
\plotone{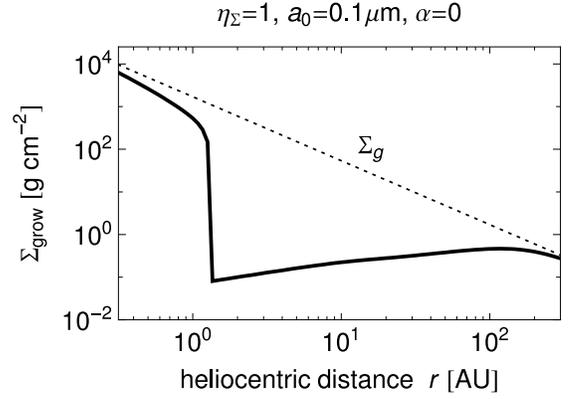}
\caption{
Gas column density $\Sigma_{\rm grow}$ of the dust growth zones
at different heliocentric distances $r$ for the fiducial model.
The dotted line show the total gas column density $\Sigma_g$.
}
\label{fig:Sigma_grow0}
\end{figure}
It will be useful to show what amount of dust 
is allowed to grow at different heliocentric distances. 
Figure~\ref{fig:Sigma_grow0} plots the column density $\Sigma_{\rm grow}$ 
of gas within the growth zones  (bimodal and unimodal zones) as a function of $r$ for the fiducial model.
Note that the column density of dust within the growth zones is $f_{dg}\Sigma_{\rm grow}$, 
while the column density of gas within the frozen zone is $\Sigma_g - \Sigma_{\rm grow}$.
For comparison, the total gas surface density $\Sigma_g$ is also shown by the dotted line.
We see that $\Sigma_{\rm grow}$ is comparable to $\Sigma_g$ at $r \la 1~\AU$
because of the presence of a large bimodal zone (see Figure~\ref{fig:rz}).
This means that the electrostatic barrier 
does not strongly affect dust growth at $r \la 1~\AU$.
Farther out the disk, however, $\Sigma_{\rm grow}$ 
steeply declines to $10^{-4}\Sigma_g \approx 10^{-1} {\rm ~g~cm^{-3}}$.
The residual value of $\Sigma_{\rm grow}$ comes from the unimodal zone at high altitudes.
At 1--10 AU, the amount of dust within the growth zone is less than $1\%$ of the total dust mass.

Figure~\ref{fig:Sigma_grow0} implies that the column density 
of the unimodal zone depends on $r$ only weakly.
As shown below, this comes from the weak radial dependence 
of the gas temperature $T$.
First, we note that the gas column density above altitude $z$, 
 $\Sigma_g^+(z)$ (Equation~\eqref{eq:Sigma+}), approximately behaves as
\beq
\Sigma_g^+(z) \approx \frac{\Sigma_g}{\sqrt{2\pi}}\frac{H}{z} {\rm e}^{-z^2/2H^2 }
\label{eq:Sigma+_approx}
\eeq
at $z \gg H$ (this equation directly follows from the asymptotic expansion of the 
complementary error function ${\rm erfc}(x)$).
Using this approximation, Equation~\eqref{eq:EEND_out} can be rewritten as 
\beq
{\cal E}_E(N_D) 
\approx
35\pfrac{a_0}{0.1~\micron}^{-3/2}\pfrac{T}{100{~\rm K}}\pfrac{\Sigma_g^+}{1~{\rm g~cm^{-2}}}.
\label{eq:EEND_approx}
\eeq
This equation means that, for fixed $r$, the column density above a given altitude $z$ 
is proportional to ${\cal E}_E(N_D)$ at that altitude.
Substituting ${\cal E}_E(N_D) = 6$ and $\Sigma_g^+ = \Sigma_{\rm grow}/2$ into 
this equation,
we obtain the column density $\Sigma_{\rm grow}$ of the unimodal growth region as
\beq
\Sigma_{\rm grow} \approx 0.3
\pfrac{a_0}{0.1~\micron}^{3/2}\pfrac{T}{100{~\rm K}}^{-1} {\rm g~cm^{-2}}.
\label{eq:Sigma_grow_approx}
\eeq
Thus, we find that the column density of the unimodal zone is inversely proportional to
$T$ and hence depends on $r$ only weakly.
It should be noted that the column density of the unimodal zone is independent of 
$\zeta$, $\Sigma_g$, and $f_{dg}$.

\subsection{Dependence on the Model Parameters}
\subsubsection{Disk Mass}
\begin{figure}
\plotone{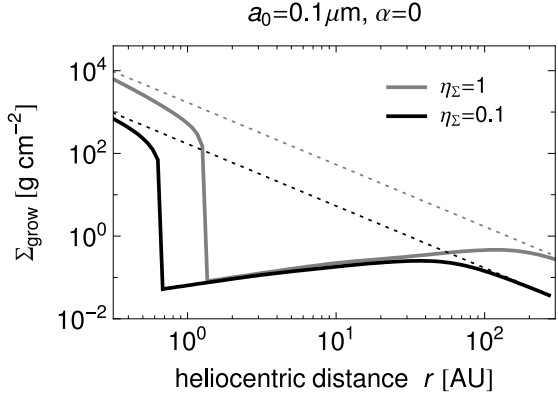}
\caption{
Same as Figure~\ref{fig:Sigma_grow0}, but for different values of disk mass.
The black solid and dotted curves indicate $\Sigma_{\rm grow}$ and 
$\Sigma_g$ for $\eta_\Sigma = 0.1$, respectively.
The gray curves are for the fiducial model of $\eta_\Sigma = 1$.
}
\label{fig:Sigma_grow_Sigma}
\end{figure}

To examine the effect of the disk mass, in Figure~\ref{fig:Sigma_grow_Sigma},
we plot the radial profile of $\Sigma_{\rm grow}$ for a lighter disk model of $\eta_\Sigma = 0.1$.
As in the fiducial case, the high value of $\Sigma_{\rm grow}$ at small $r$ corresponds 
to the bimodal zone, while the floor of $\Sigma_{\rm grow}$ at larger $r$
comes from the unimodal zone.
We find that the bimodal  zone shrinks as the disk mass decreases. 
This is because ionization sources from upper layers (cosmic rays and X-rays)
can penetrate to the midplane more easily when the disk mass is smaller.
However, this effect is limited because a smaller disk mass also causes
a higher settling velocity and hence a higher collisional velocity.
Moreover, the column density of the unimodal zone is very insensitive to $\eta_\Sigma$
as is already expected from Equation~\eqref{eq:Sigma_grow_approx}.
Thus, we can conclude that  the change in the disk mass 
has only a minor effect on the size of the frozen zone.

\subsubsection{Monomer Size}
We have assumed that dust growth begin with single-sized monomers.
In reality, however, dust monomers in protoplanetary disks  will obey a certain size distribution.
For instance, interstellar extinction implies that the grain sizes
range from $0.005~\micron$ to $0.25~\micron$ \citep{MRN77}.
To fully take into account the monomer size distribution is challenging
since the porosity model for aggregates would be much more complicated.
For this reason, we try to estimate the effect of monomer size distribution 
by varying the monomer size $a_0$ in our model.  
\begin{figure}
\plotone{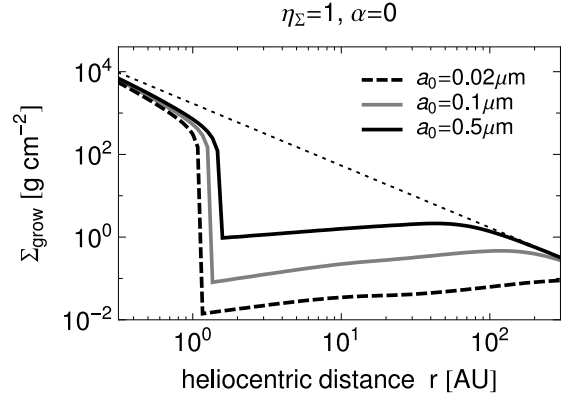}
\caption{
Same as Figure~\ref{fig:Sigma_grow0}, but for different monomer radii $a_0$.
The black solid and dashed curves show $\Sigma_{\rm grow}$ 
for $a_0 = 0.5~\micron$ and $0.02~\micron$, respectively.
The gray curve is for the fiducial model of $a_0 = 0.1~\micron$.
}
\label{fig:Sigma_grow_a0}
\end{figure}

Figure \ref{fig:Sigma_grow_a0} shows $\Sigma_{\rm grow}$ 
for three different monomer radii, $a_0 = 0.02$, $0.1$, and $0.5~\micron$.
We see that the increase of $a_0$ by a factor of 5
leads to the enhancement of the column density of the unimodal zone 
(i.e., $\Sigma_{\rm grow}$ at $r\ga 1~\AU$) by a factor of $10$.
This is consistent with Equation~\eqref{eq:Sigma_grow_approx} predicting that
the column density scales as $a_0^{3/2}$.
The growth zone increases with $a_0$ 
because aggregates made of larger monomers have 
larger mass-to-area ratios and hence higher drift velocities.
However, even with $a_0=0.5~\micron$ (which is twice the maximum size in the MRN distribution),
the enhanced $\Sigma_{\rm grow}$ is still much smaller
than the total surface density $\Sigma_g$.
Indeed, $a_0$ must be as large as $10~\micron$ 
for $\Sigma_{\rm grow}$ to be comparable to $\Sigma_g$ at every $r$.
Thus,  a change of the monomer size within a realistic range gives only a minor effect.

\subsubsection{Turbulence}
\begin{figure}
\plotone{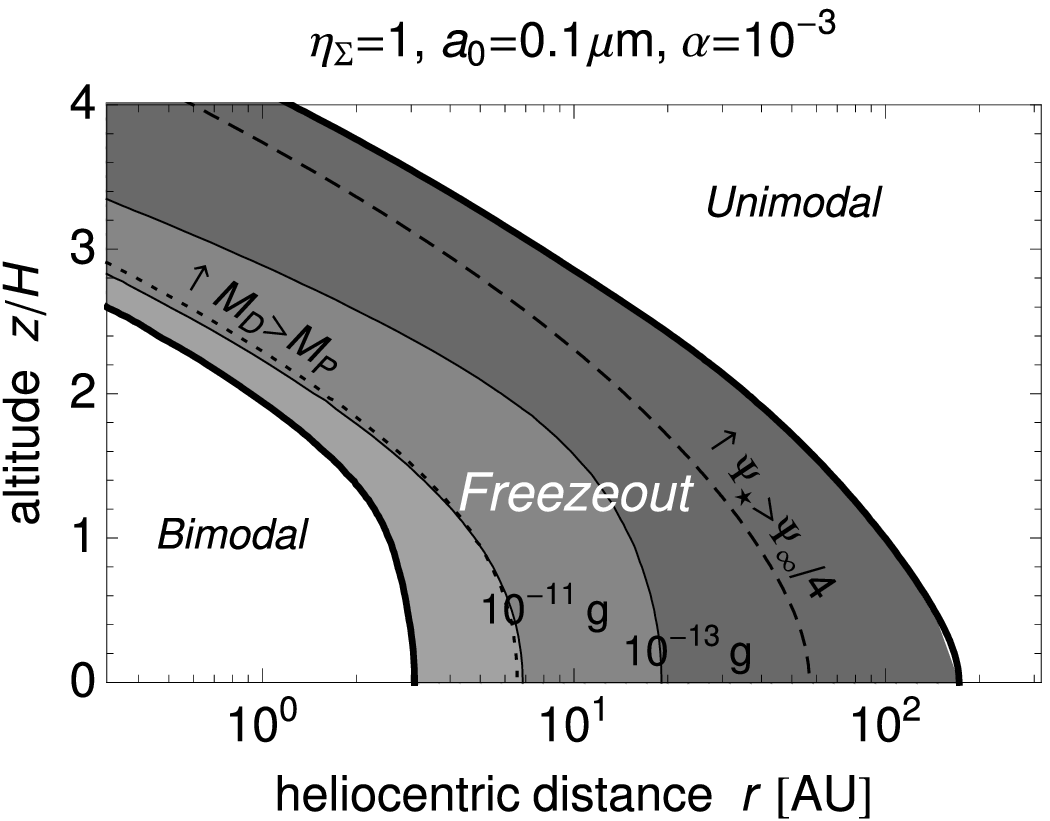}
\plotone{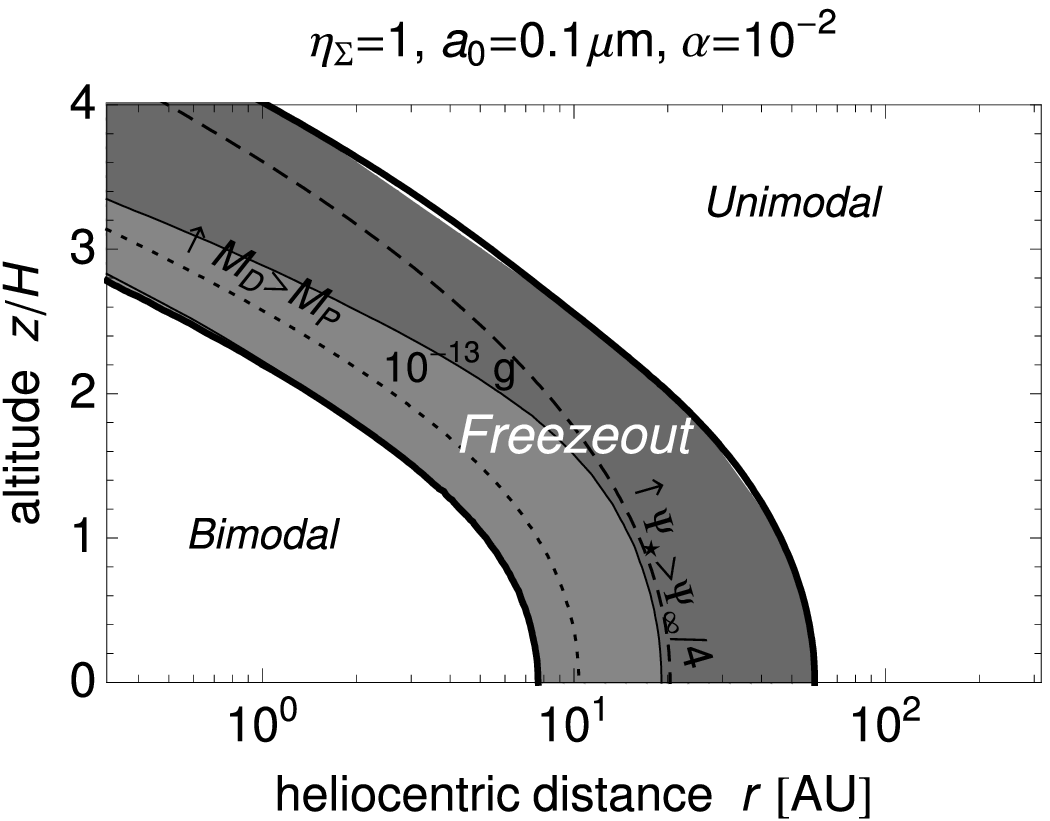}
\caption{
Same as Figure~\ref{fig:rz}, but for different turbulence strengths $\alpha$.
The upper and lower panels are for $\alpha = 10^{-3}$ and $10^{-2}$, respectively.
}
\label{fig:rz_alpha}
\end{figure}

\begin{figure}
\plotone{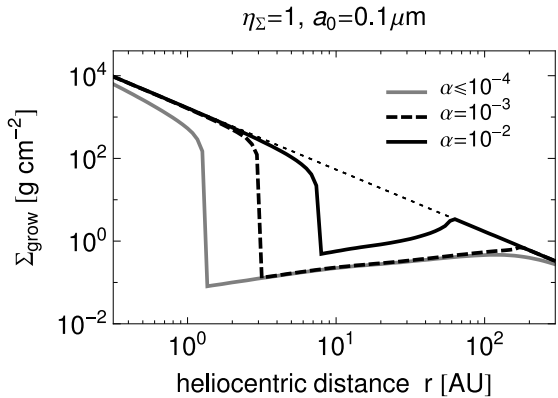}
\caption{
Same as Figure~\ref{fig:Sigma_grow0}, butfor different turbulence strengths $\alpha$.
The  black solid and dashed curves correspond to $\alpha = 10^{-2}$ and $10^{-3}$, respectively.
The gray curve is for even weaker turbulence, $\alpha \leq 10^{-4}$.
}
\label{fig:Sigma_grow_alpha}
\end{figure}
Turbulence enhances the collision velocity of aggregates,
so it acts to reduce the size of the frozen zone.
To see this effect, we have computed $\Sigma_{\rm grow}$ with changing the value of $\alpha$.
The two panels in Figure~\ref{fig:rz_alpha} show
 the two-dimensional maps of the frozen zone for $\alpha = 10^{-3}$ and $10^{-2}$, 
and the radial profiles of $\Sigma_{\rm grow}$ for the two cases 
are plotted in Figure~\ref{fig:Sigma_grow_alpha}.
We have also computed $\Sigma_{\rm grow}$ for $\alpha \leq 10^{-4}$, 
but the result is indistinguishable from that for the fiducial laminar case.
Comparing Figure~\ref{fig:rz_alpha} with Figure~\ref{fig:rz},
we find that the presence of turbulence is particularly important near the midplane.
This is because the settling velocity vanishes at the midplane while the turbulence-driven velocity does not.
For $\alpha = 10^{-3}$ and $10^{-2}$,
the bimodal zone expands out to $3~\AU$ and $10~\AU$, respectively. 
This means that strong turbulence with $\alpha \sim 10^{-2}$ removes
the frozen zone from planet-forming regions. 
At larger $r$, however, removal of the frozen zone requires even higher values of $\alpha$.
We find that  $\alpha$ must be higher than $10^{-1}$ 
for $\Sigma_{\rm grow}$ to be comparable to $\Sigma_g$ at every $r$.

Unfortunately, it is uncertain what is a realistic value of $\alpha$ for protoplanetary disks,
especially in early stages of dust evolution.
It is commonly believed that the most promising mechanism driving disk turbulence is 
the magnetorotational instability \citep[MRI;][]{BH91}.
MHD simulations suggest that the MRI can sustain turbulence 
at a level of $\alpha \sim 10^{-2}$ or higher 
depending on the net vertical flux of the magnetic field \citep{SMI10}. 
This implies that the MRI could assist dust to overcome  
the electrostatic barrier at $r\la 10~\AU$ (see Figure~\ref{fig:Sigma_grow_alpha}).
The problem is whether the MRI does operate there in early stages of dust evolution.
Since the MRI is an MHD phenomenon, 
it requires an ionization degree high enough for the gas to couple to the magnetic fields.
The MRI is suppressed by Ohmic dissipation
where the ionization degree is too low, which is known as the ``dead'' zone.
Importantly, the size of the dead zone can be very large when 
small and fluffy dust aggregates are abundant 
because they have large surface areas and therefore efficiently capture the ionized gas.
For example, if all dust particles are  $0.1\micron$ in size,
the dead zone can extend to $20~\AU$ \citep{Sano+00}.
Furthermore, the large dead zone remains while the particles grow into fractal aggregates 
because their total projected area is nearly conserved \citepalias{O09}.
These facts suggest that MRI-driven turbulence may be absent from inside $10$ AU 
in the early stage of dust evolution. 
However, the physics determining the size of the dead zone 
is not yet fully understood \citep[see, e.g.,][]{IS05}, 
so further investigation is needed on this issue. 

\section{Discussion}
\subsection{On the Role of Porosity Evolution}
So far, we have assumed that dust grows into fractal aggregates of $D\sim 2$.
The assumption of fractal evolution is reasonable when analyzing the electrostatic barrier 
since the freezeout occurs much earlier than the onset of collisional compaction (see Section 4.1).
However, one cannot rule out that some unknown process could cause compaction of aggregates.
Therefore, it will be useful to see how the result in the previous section 
depends on the porosity model adopted.

Here, we consider the conventional, {\it compact} growth model where 
aggregates grow as zero-porosity spheres \citep[e.g.,][]{NNH81,THI05,BDH08}.
The effect of the electrostatic barrier on the compact dust growth has been already 
investigated in our previous paper \citepalias{OTTS11a}.
We have found that the freezeout criterion for the compact model 
can be again written as Equation~\eqref{eq:growthcond},
but, the electrostatic energy ${\cal E}_E(N_D)$ is now given by
 \citepalias[see Equation~(66) of][]{OTTS11a}
\beq
{\cal E}_E(N_D) = \frac{f_E}{2} \left[1+(hf_D^{-3/5})^{-0.8}\right]^{-2.5}f_D^{-1/5},
\label{eq:EEND_comp}
\eeq 
where the definitions of $f_D$, $f_E$, and $h$ are the same as those for the porous case
(i.e., Equations~\eqref{eq:fD}, \eqref{eq:fE}, and \eqref{eq:h}, respectively).
It can be easily checked from Equations~\eqref{eq:fDSr}--\eqref{eq:hr}
that the right-hand side of Equation~\eqref{eq:EEND_comp} is independent of the monomer radius $a_0$.
This should be so since ``monomers'' are not well defined in the compact model.

\begin{figure}
\plotone{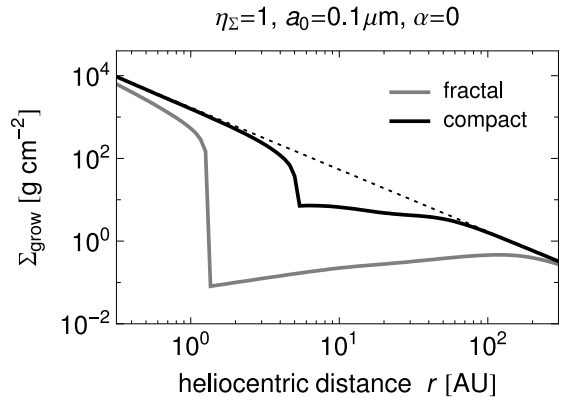}
\caption{
Gas column density $\Sigma_{\rm grow}$ of the growth zones
as a function of $r$ for the {\it compact} growth model (solid curve).
For comparison, the result for the fiducial porous model is indicated by the grey curve.
The dotted line show the total gas column density $\Sigma_g$.
}
\label{fig:Sigma_growc}
\end{figure}
Using the freezeout criterion and Equation~\eqref{eq:EEND_comp},
we can compute the column density $\Sigma_{\rm grow}$ of the growth zone 
in the same way as we did in Section 4.
Figure \ref{fig:Sigma_growc} plots $\Sigma_{\rm grow}$ for the compact dust model
with the fiducial parameter $\eta_\Sigma = 1$ and $\alpha = 0$.
For comparison, we overplot the result for the fiducial porous model
shown in Figure~\ref{fig:Sigma_grow0}.
We see that the frozen zone predicted by the compact model 
is considerable smaller than that by the porous model.
This reflects the fact that the differential drift velocity of compact ``aggregates'' 
grows more rapidly with $N$ and hence 
takes over the thermal velocity at lower $N$ (see \citetalias{OTTS11a}).

From the above example, we can say that 
the size of the frozen zone is quite sensitive to porosity evolution.
However, it should be kept in mind that evolution into highly porous aggregates are inevitable 
unless some compaction mechanism exists (e.g., \citealt*{WB98,Blum+98,KPH99}; \citetalias{OTS09})
and that collisional compaction is not the mechanism in the early stage of dust evolution. 
Of course, there remains a possibility that there exist any other effective mechanisms.
For example, compaction of fluffy aggregates is in principle possible 
if a number of small particles are supplied by some process and fill in the voids in the aggregates.
To specify such mechanisms is beyond the scope of this paper, so we leave this issue open for future work.

\subsection{Effects of Global Dust Transport}
In Section~4.2.3, we have seen that turbulence as strong as $\alpha \ga 10^{-2}$ 
is preferable for dust growth beyond the electrostatic barrier.
However, from the viewpoint of planetesimal formation, 
strong turbulence is not always preferable because it causes 
the fragmentation barrier against large and compacted aggregates.
For example, large and compacted aggregates with $\tau_f \sim \Omega_{\rm K}^{-1}$
acquires a random velocity of $ \Delta u \approx u_L \approx \sqrt{\alpha} c_s$  
in turbulence \citep[e.g.,][]{OC07}.
If we put $\alpha \ga 10^{-2}$ and $c_s \approx 700~{\rm m~s^{-1}}$ 
(as is for $T \approx 100~{\rm K}$),
the random velocity amounts to $\Delta u\ga 70~{\rm m~s^{-1}}$. 
By contrast, it is suggested from computer and laboratory collision experiments
that the threshold collision velocity for catastrophic disruption to occur is
$\approx 50~{\rm m~s^{-1}}$ for icy aggregates \citep{Wada+09} 
and is an order of magnitude even lower for rocky aggregates \citep{BW08,Guettler+10}.
Thus, even if we assume a high threshold velocity of $\approx 50~{\rm m~s^{-1}}$,
the large aggregates cannot grow beyond the fragmentation barrier.

For this reason, it is worth considering whether planetesimal formation is possible
in the frozen zone {\it without} strong turbulence.
It is clearly difficult to form planetesimals directly from fluffy frozen aggregates,
because fluffy aggregates couple to the gas too strongly to   
form gravitationally unstable dust layers or clumps.
Here, we examine a possibility that 
large and compacted aggregates are supplied from {\it outside} the frozen zone
through some dust transport mechanism.

For such transport mechanisms, we focus on
(1) vertical mixing of frozen aggregates by (weak) turbulence
and (2) radial infall of aggregates grown at outer ($\ga 100$ AU) regions.
As shown below, both mechanisms can supply large and compact aggregates 
to the frozen zone on a timescale of $10^{6}$ yr or longer.

\begin{figure}
\epsscale{1.1}
\plotone{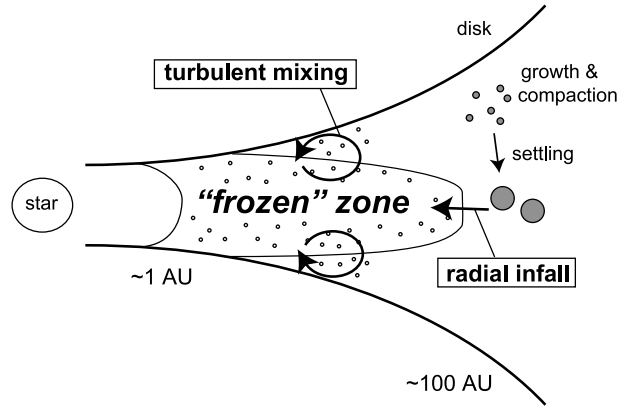}
\caption{
Dust transport mechanisms that could supply 
large aggregates to the frozen zone (not to scale).
Without any transport mechanism, 
the electrostatic barrier halts dust growth at the fractal growth stage
at $1$--$10~\AU \la r \la 100~\AU$ near the midplane (the ``frozen'' zone).
However, vertical mixing due to turbulence could allow
the frozen aggregates to grow outside the frozen zone. 
Furthermore, already large and compacted aggregates can be supplied from 
outer ($\ga 100~\AU$) regions to the frozen zone as a result of their radial infall.
Both processes can provide large and compacted aggregates on a timescale of  $\ga 10^6~{\rm yr}$
(see text).
}
\label{fig:outside-in}
\end{figure}

\subsubsection{Vertical Mixing by Weak Turbulence}
As seen in the previous sections, turbulence with $\alpha \la 10^{-4}$
does not significantly contribute to the collision velocity of aggregates.
However, turbulence does not contribute only to the collisional velocity but also 
to the vertical mixing of dust material.
Vertical mixing is efficient even by weak turbulence
because fluffy aggregates strongly couple to the gas.
The mixing will allow the frozen aggregates to go out of the frozen zone and grow there
until reentering the frozen zone (see Figure~\ref{fig:outside-in}).
This cycle can prevent the aggregates from being perfectly frozen. 

To fully examine this effect, we would have to solve the coagulation equation including vertical diffusion
\citep[as done by, e.g.,][]{DD05}, but this is beyond the scope of the present study.
In this paper, we restrict ourselves to simply estimating 
how long time is required for dust in the frozen zone
to grow beyond the fractal growth regime ($M > M_{\rm roll}$)
with the help of the turbulent diffusion. 
To do so, we estimate the mean growth time ($\approx$ mean collision time) of aggregates at fixed $r$,
\beq
\tau_{\rm grow} \equiv \left(\frac{1}{M}\frac{dM}{dt} \right)^{-1},
\label{eq:t_grow_def}
\eeq
where $M$ is the mass of the aggregate.
If the electrostatic barrier were absent, we would have 
$dM/dt = \rho_d \sigma_{\rm coll} \Delta u$,
and hence $\tau_{\rm grow} = M/\rho_d \sigma_{\rm coll} \Delta u$,
where $\rho_d$ is the dust density, $\Delta u$ is the collision velocity,  
and $\sigma_{\rm coll}$ is the collisional cross section.
Here, we need to take into account the fact that 
aggregate collision is only allowed in the growth zone 
with the gas column density $\Sigma_{\rm grow}$.
Assuming that the dust is vertically well mixed by turbulence, 
the probability that an aggregate experiences a collision in the growth zone 
will be given by the ratio $\Sigma_{g}/\Sigma_{\rm grow}$.
Therefore, we evaluate $\tau_{\rm grow}$ as
\beq
\tau_{\rm grow} = \frac{M}{\rho_d \sigma_{\rm coll} \Delta u}\frac{\Sigma_g}{\Sigma_{\rm grow}}.
\label{eq:t_grow}
\eeq
To simplify Equation~\eqref{eq:t_grow}, 
we use the fact that aggregates grow mainly through 
similar-sized collisions \citepalias{OTS09}
and approximate $\sigma_{\rm coll}$ as $4\pi a^2$.
A collision takes the longest time when the relative velocity $\Delta u$ minimizes.
Neglecting the contribution of weak turbulence to $\Delta u$, 
the minimum value of $\Delta u$ is evaluated as the differential settling velocity,
$\Delta u \approx \eps g \tau_f \approx \eps \Omega_{\rm K}^2 H \tau_f$.
Using $\tau_f \approx M/\Sigma_g\Omega_{\rm K}A$, $A \approx \pi a^2$, 
and $\rho_d \sim f_{dg}\Sigma_g/H$, we obtain
\beqn
\tau_{\rm grow} &\approx& \frac{1}{4\eps f_{dg} \Omega_{\rm K}}
\frac{\Sigma_g}{\Sigma_{\rm grow}}
\nonumber \\
&\approx& 4 \times 10^4
 \pfrac{10^{-1}}{\eps}\pfrac{r}{100~\AU}^{3/2}\frac{\Sigma_g}{\Sigma_{\rm grow}} ~{\rm yr}.
\label{eq:t_grow_approx}
\eeqn
Note that the right-hand side of Equation~\eqref{eq:t_grow_approx} does not involve $M$.

\begin{figure}
\plotone{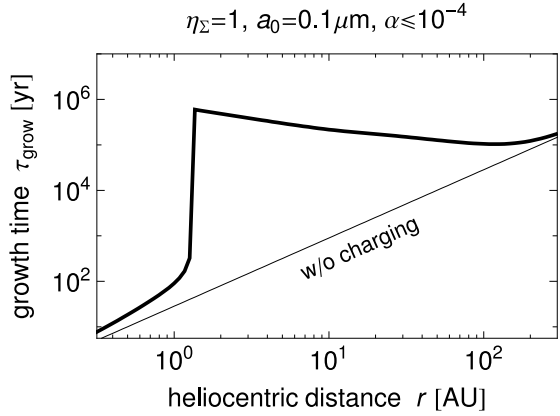}
\caption{
Timescale of dust growth driven by turbulent diffusion, 
 $\tau_{\rm grow}$ (Equation~\eqref{eq:t_grow_approx}; thick solid curve).
The turbulence parameter is set to  $\alpha \leq 10^{-4}$.
The thin solid curve shows $\tau_{\rm grow}$ when the frozen zone is absent.
The parameters $\eta_\Sigma$ and $a_0$ are the same as those for the fiducial model. 
}
\label{fig:rate}
\end{figure}
In Figure~\ref{fig:rate}, we plot $\tau_{\rm grow}$ as a function of $r$ 
for a weakly turbulent case of $\alpha \leq 10^{-4}$.
The jump in $\tau_{\rm grow}$ at $r \approx 1~\AU$
corresponds to the outer edge of the bimodal growth zone (see Figure~\ref{fig:Sigma_grow_alpha}).
At $1~\AU \la r \la 100~\AU$, where $\Sigma_{\rm grow} \ll \Sigma_g$,  
the typical value of $\tau_{\rm grow}$  is about $10^{5}$ yr (at 100~AU) 
to $10^{6}~{\rm yr}$ (at 1~AU).

Note that $\tau_{\rm grow}$ only represents 
the mean time spent for a single collision.
What we really want to know is the time required to grow beyond the fractal growth regime.
Using the fact that $\tau_{\rm grow} = dt/d\ln M$ is independent of $M$ at $M>M_D$
and is shorter at $M<M_D$, the required time is estimated as  
$\tau_{\rm grow}\ln(M_{\rm roll}/M_D)$.
As found from the lower panel of Figure~\ref{fig:EN0}, 
this time is longer than $\tau_{\rm grow}$ by a factor of $\sim 10$.
Hence, we conclude that turbulent mixing could assist the frozen aggregates
to grow beyond the fractal growth regime on a timescale of $\ga10^6$ yr.
 
It is interesting to note that the vertical mixing predicts a quite flat radial profile 
for the dust growth timescale at $1~\AU \la r \la 100~\AU$.
In fact, it can be directly shown from  Equations~\eqref{eq:Sigma_grow_approx} and 
\eqref{eq:t_grow_approx} that the growth time at the locations is proportional to
 $\Sigma_g/\Omega_{\rm K}\Sigma_{\rm grow} \propto T \Sigma_g/\Omega_{\rm K}$
and hence scales with $r^{-1/2}$ in the MMSN model\footnote{$\tau_{\rm grow}$ is 
{\it independent} of $r$ if $\Sigma_g \propto r^{-1}$ as suggested 
by submillimeter observations \citep[e.g.,][]{AW07,Andrews+09} 
and constant-$\alpha$ accretion disk models \citep[e.g.,][]{Hartmann+98}.
}.
The flat profile is in marked contrast to previous uncharged dust models 
where $\tau_{\rm grow}$ scales with the orbital period $(\Omega_{\rm K}^{-1})$ 
and is hence as steep as $r^{3/2}$.

\subsubsection{Radial Infall of Large Aggregates from Outer Regions}
Aggregates outside the frozen zone are allowed to
grow beyond  $M = M_{\rm roll}$ and experience collisional compaction.
As the compaction proceeds, they will gradually decouple from the gas,
settle onto the midplane, and drift towards the central star \citep{W77}.
Aggregates falling from outer ($r \ga 100~\AU$) growth zones  
should enter the frozen zone (see Figure~\ref{fig:outside-in}).
The time spent for the onset of the radial drift will be
comparable to that for the onset of the collisional compaction,
which is $\sim 10 \tau_{\rm grow} \sim 10^6$ yr 
at $r\approx 100~\AU$ (see Section~5.2.1).
Thus, large and compact aggregates can be supplied from the outer regions 
on a timescale of $\sim 10^6$ yr. 

Furthermore, the drifting aggregates have a sufficiently high kinetic energy 
to sweep up frozen ones beyond the electrostatic barrier.
To show this, we compare the impact energy $E_K$ with the electrostatic energy $E_E$ for the collision.
The lower limit of $E_K$ can be estimated by considering the lower 
limit of the radial drift velocity $u_r$ for the drifting aggregates.
Here, it is useful to note that the timescale of the radial infall is given by 
\beq
\tau_{\rm infall} \equiv \frac{r}{|u_r|} 
\approx 10^5 \pfrac{r}{100~\AU}\pfrac{|u_r|}{5~{\rm m~s^{-1}}}^{-1}~{\rm yr}.
\label{eq:t_infall}
\eeq
Dust aggregates continue growing at the same location until $\tau_{\rm infall}$ 
becomes shorter than the local growth timescale $\tau_{\rm grow}$ (see Section~5.2.1 for its definition).
Hence, the lower limit of $u_r$ for the drifting aggregates can be evaluated 
from the inequality $\tau_{\rm infall}<\tau_{\rm grow}$. 
At $r \approx 100~\AU$, this inequality gives $u_r \ga 5~{\rm m~s^{-1}}$.
Thus, the lower limit of $E_{\rm K}$ is estimated as
\beq
E_K \ga \frac{1}{2}M_Fu_r^2 \ga 10^{-9} \pfrac{M_F}{10^{-14}{\rm~g}}~{\rm erg},
\eeq
where we have used the fact that the mass $M_F$ of the frozen aggregates 
is much smaller than that of the drifting aggregates.
For the electrostatic energy, we can obtain its upper limit by setting $Q \leq \Psi_\infty a\kB T/e$.
Assuming that the frozen aggregates is much smaller than the drifting aggregates,  
we obtain 
\beqn
E_E &\la& a_F \pfrac{\Psi_\infty \kB T}{e}^2  \nonumber \\
&\sim& 10^{-14} \pfrac{M_F}{10^{-14}{\rm~g}}^{1/2}\pfrac{T}{30{\rm~K}}^2~{\rm erg},
\eeqn
where $a_F = a_0 (M_F/m_0)^{1/2}$ is the radius of the frozen aggregates.
Since $E_K$ is much higher than $E_E$, 
we conclude that  the radially drifting aggregates can overcome the electrostatic barrier 
to collide with the frozen aggregates.

However, it is unclear from the above argument 
whether the collision results in the growth or fragmentation (erosion) of the drifting aggregates.
Interestingly, recent laboratory experiments \citep{TW09,Guettler+10} 
show that net growth of a large and compact aggregate is possible
even at high collision velocities ($> 1~{\rm m~s^{-1}}$ for rocky aggregates)
if the projectiles are much smaller than the target
(see the {\it``pC''}-regime in Figure~11 of \citealt{Guettler+10}).
Hence, the presence of frozen aggregates could even assist 
the growth of the drifting aggregates beyond the fragmentation barrier.

\subsubsection{Implications for the Timescale of Protoplanetary Dust Growth}
As discussed above, 
dust evolution in the frozen zone is possible on a timescale of $10^6~{\rm yr}$ or longer.
By contrast, if the electrostatic barrier is absent, the collision timescale for fractal aggregates 
at $r \la 10~\AU$ is as short as  $\la 10^3$ yr (see Figure~\ref{fig:rate}).
Thus, the electrostatic barrier can dramatically alter the timescale 
of dust evolution in protoplanetary disks.

There is possible evidence that evolution 
of small grains/aggregates occurs on a long timescale.
Infrared observations of classical T Tauri stars 
show mid-infrared excess evolving on a timescale of $10^6$ yr \citep[e.g.,][]{Furlan+06}. 
This is usually interpreted as an indication that micron-sized and warm ($\ga 100~{\rm K}$) 
grains grow in their curcumstellar disks on this timescale \citep[e.g.,][]{DD05}.
However, it is also possible to interpret this timescale as the duration of fractal dust growth
because the spectral signature of large and fractal ($D\sim 2$) aggregates 
is similar to that of small and compact particles \citep{Min+06}.
Interestingly, the presence of the frozen zone at $r\ga 1~\AU$ 
together with the dust transport mechanisms considered above
can explain the retention of fractal aggregates in warm regions ($\la 10~\AU$) 
on a timescale $\sim 10^6~{\rm yr}$.
Previously, the retention of fast-growing small grains has been attributed 
to the collisional destruction of large aggregates into fragments \citep[e.g.,][]{DD05,DD08,BDB09}.
However, the electrostatic barrier can be an alternative solution to this problem. 

Another piece of possible evidence for slow dust evolution may be obtained from primitive meteorites.
Radioisotope dating of chondrites has revealed that the the formation of chondrules 
began 1--3 million years after the formation of Ca-Al-rich inclusions  \citep[e.g.][]{Kita+00,Kita+05}.
This implies that the growth of chondrule precursors (i.e., mm-sized dust aggregates) 
occurred on a timescale of $10^6$ yr.
The electrostatic barrier or the ``bouncing barrier'' proposed by \citet{Zsom+10}
might have played a role in this slow accretion process.

\section{Summary}
In this paper, we have examined where in a protoplanetary disk
dust charging can halt local dust evolution.  
This is the first step towards the modeling of dust evolution in protoplanetary disks
including dust charging together with other important mechanisms, such as collisional compaction
\citep{SWT08}, radial drift \citep{W77,BDH08}, bouncing \citep{Zsom+10}, 
and fragmentation \citep{BDH08,BDB09}. 
Our findings are summarized as follows.

\begin{enumerate}
\item
We find a ``frozen zone'' where dust growth stalls in the fractal growth stage.
For weakly turbulent disks ($\alpha \la 10^{-2}$), 
the frozen zone contains the major part of dust materials at a few AU to 100 AU from the central star.
Dust growth beyond the fractal stage is only allowed 
in an inner region where ionizing cosmic rays and X-rays do not reach,
and in an outer region where settling velocity is high enough to overcome the electrostatic barrier.
The freezeout mass, the mass at which the growth begins to stall,
strongly depends on the distance from the central star,
typically ranging from $10^{-7}~{\rm g}$ (at $\sim$ 1 AU) 
to as small as $10^{-13}~{\rm g}$ (at $\ga$ 10 AU).

\item 
The size of the frozen zone does not significantly change with changing
the disk mass and monomer size within a realistic range (Sections 4.1 and 4.2).
By contrast, turbulence as strong as $\alpha \ga 10^{-2}$
can help dust in the ``planet-forming'' region ($\la$10~AU)
to overcome the electrostatic barrier (Section 4.3).
Caution is needed when considering this result in the context of planetesimal formation,
since such strong turbulence can cause the fragmentation barrier {\it after} the fractal growth stage.

\item For weakly turbulent disks, 
we have considered two dust transport mechanisms 
that could lead to dust evolution in the frozen zone (Sections~5.2).
Turbulent mixing across the boundary of the frozen zone 
can prevent dust growth from being completely frozen.
Large and compacted aggregates can be also supplied from large 
heliocentric distances  ($\ga$ 100 AU) through their radial infall.
Both mechanisms can result in the supply of large and compacted aggregates
and the removal of small and fractal aggregates on a timescale of $10^6$ yr or longer.
This is in clear contrast to previous theoretical understanding that without fragmentation,
small dust particles get depleted in disks on much shorter timescales \citep[e.g.,][]{DD05,DD08}. 
This might explain the ``slow'' ($\sim 10^6$ yr) dust evolution 
suggested by infrared observation of T Tauri stars \citep[e.g.,][]{Furlan+06}
and by radioactive dating of chondrules \citep[e.g.,][]{Kita+05}.
\end{enumerate}

Finally, we remark that the above findings are obtained 
using a fractal growth model for small dust aggregates.
This is in marked contrast to most previous studies on dust coagulation
where aggregates are simplified as compact spheres \citep[e.g.,][]{NNH81,THI05,BDH08}.
However, it is critical to properly take into account porosity evolution because 
compact spheres would be more resistive to the electrostatic barrier (Section 5.1).
On the other hand, it is also true that the effect of the electrostatic repulsion could 
be reduced if some mechanism prevents aggregates from being highly fluffy ($D\sim 2$).
As shown in Section~4.1, collisional compaction is very unlikely to be the mechanism
because the freezeout begins much earlier than it becomes effective.
Nevertheless, we cannot rule out possible existence of any other compaction mechanisms.
We leave this issue open for future work.

\acknowledgments
We thank the anonymous referee for careful reading 
of the manuscript and useful comments that helped improve it.
S.O. is supported by Grants-in-Aid for JSPS Fellows ($22\cdot 7006$) from MEXT of Japan.



\end{document}